\documentstyle[epsfig,natbib2,natbibmnfix]{mn}

\newcommand{\be}{\begin{equation}}
\newcommand{\ee}{\end{equation}}
\newcommand{\bea}{\begin{eqnarray}}
\newcommand{\eea}{\end{eqnarray}}
\newcommand{\bc}{\begin{center}}
\newcommand{\ec}{\end{center}}
\newcommand{\ol}[1]{ {\overline{#1}}}
\newcommand{\lu}{\,h^{-1}{\rm kpc}}

\newcommand{\dd}{{\rm d}}

\newcommand{\tree}{{\small TREE}}
\newcommand{\grape}{{\small GRAPE}}
\newcommand{\gadget}{{\small GADGET}}

\bibpunct[,]{(}{)}{;}{a}{}{,}

\title{Tidal tails in CDM cosmologies}

\author[V.\ Springel and S.\ D.\ M.\ White]
{Volker Springel and Simon D. M. White \\
Max-Planck-Institut f\"{u}r Astrophysik, 
Karl-Schwarzschild-Stra\ss{}e 1, 
85740 Garching bei M\"{u}nchen, Germany}

\begin{document}

\maketitle
\begin{abstract}
We study the formation of tidal tails in pairs of merging 
disk galaxies with structural properties motivated by current theories
of cold dark matter (CDM) cosmologies. 
In a recent study, 
\citet*{Du96}
showed that 
the formation of prominent tidal tails can be strongly suppressed 
by massive and extended dark haloes.
For the large halo-to-disk mass ratio expected in CDM cosmologies
their sequence of models failed to produce strong tails like those
observed in many well-known pairs of interacting galaxies.
In order to test whether this effect can constrain the viability of CDM
cosmologies, 
we construct N-body models of disk galaxies with structural
properties derived in analogy to  
the analytical work of \citet*{Mo98}.
With a series of self-consistent collisionless simulations of
galaxy-galaxy mergers we demonstrate
that even the disks of
very massive dark haloes have no problems developing long tidal
tails, provided the halo spin parameter is large enough.
We show that the halo-to-disk mass ratio is a poor indicator for
the ability to produce tails. Instead, the relative size of disk and halo, or
alternatively, 
the ratio of circular velocity to local escape speed at the 
half mass radius of the disk
are more useful criteria.
This result holds 
in all CDM cosmologies. 
The length of tidal tails is thus unlikely
to provide useful constraints on such models.
\end{abstract}

\section{Introduction}

In standard hierarchical scenarios for galaxy formation, mergers of
galaxies are common events that lead to the build-up of ever more
massive galaxies. In fact,
such mergers have been observed for a long
time. 
There is now a large database of 
well studied examples of merging or strongly
interacting disk galaxies, among the most prominent of them are 
NGC4038/39 (the Antennae), NGC4676 (the Mice), and NGC7252.
Many of these pairs feature extended {\em tidal tails}, with a length
that can
reach more than 100$\lu$ in projection, or in the extreme case of IRAS19254-7245 (the
Superantennae) even $\sim 305\lu$ from tip to tip \citep{Co91}.

The tails originate in 
close encounters
of disk galaxies,
when the mutual tidal field ejects disk stars into arcing
trajectories that lead to the formation of long tails pointing way
from the galaxies, and of bridges
connecting them. 
This process was first demonstrated convincingly in a
classic paper by \citet{To72}. 
Later \citet{Wh78,Wh79} computed
the first fully self-consistent 3-dimensional simulations of merging 
galaxies and established the rapidity of the orbital decay, 
and the structural resemblence of the merger remnants to elliptical galaxies.
This work has been confirmed and extended over the years by
simulations with increasingly realistic initial conditions and ever
better numerical resolution \citep{Fa82,Fa83,Ne83,Ba88,Ba89,Ba92,He92,He93d,Ba96}.
There have
also been quite successful attempts to model particular interacting
systems in detail, for example NGC7252 \citep{Hi95} and NGC2442 \citep{Mi97}.

Recently, \citet*[hereafter DMH]{Du96} 
studied the morphology of tidal tails in a series of merging models of disk
galaxies with varying halo-to-disk mass ratio. 
In their sequence of four models,
they kept the inner
rotation curve very nearly constant and surrounded the disk and the
bulge with ever more extended and massive dark haloes.
They found that with
increasing mass of the dark halo, the resulting tidal tails
became shorter and less massive. 
Their explanation for this effect is simple. For a fixed
structure of the disk, a more massive halo leads to a deeper
potential well and a higher encounter velocity. As a consequence, the
duration and overall strength of the perturbation to the disk is smaller,
and the perturbed material cannot as easily climb out of 
the deeper potential well.

Dubinski, Mihos, and Hernquist have 
followed up this study with an analysis of
NGC7252 \citep{Mi98}, and an investigation of different dark
matter profiles with a restricted 3-body code \citep{Du97}.
Again they found that disk models with large 
halo-to-disk mass ratio were not able to produce prominent tidal
tails.
In particular, they concluded that for mass
ratios above 10:1 it should be exceedingly difficult to make tails as
long as those observed in systems like NGC7252 or NGC4038/39.
Since
the currently favoured theoretical 
values of halo-to-disk mass are considerably
higher than this, they speculated that there might be a 
conflict with cold dark
matter (CDM) cosmologies.

In this work we examine the tail-forming ability of 
realistic models of disk galaxies, where `realistic' means
that their structural properties are motivated to a large degree
by current
theories of CDM
cosmologies. We derive the structural properties of our disk galaxies 
according to the analytic model of 
\citet*[hereafter MMW]{Mo98},
and we collide pairs of these galaxies in self-consistent N-body
simulations. We adopt initial conditions for these merger
simulations 
that are 
favourable for tail formation.

We will demonstrate that the halo-to-disk mass ratio is not
a particularly useful parameter 
for characterizing ability 
to make tidal tails. 
We find that it is the relative distribution
of disk and halo material that is relevant, not the mass ratio itself.
This conclusion was reached earlier by Barnes (1997, private
communication) through analysis of a series of simulations with
halo/disk models differing both from those of DMH and from the CDM-based
models we use here.

We will show that realistic disk models
in CDM cosmologies have no problem producing long and massive
tidal tails, provided the spin parameter of their dark halo is large enough.
This result is practically independent of the adopted halo-to-disk mass
ratio.

This work is organized as follows. In Section 2 we describe 
the structural properties of our disk models, and in
Section 3 we discuss our techniques for setting up N-body representations
of these models. A description of the different simulations we
have performed is given in Section 4, while Section 5 presents the
results.
Finally, we summarize and discuss our findings in Section 6.

\section{Models of disk galaxies}

MMW have developed 
an analytical model 
for the structure of disk galaxies
embedded in
cold dark matter haloes. Their model rests on a number of
simple yet plausible assumptions, and it is very successful in
reproducing the observed properties of disk
galaxies. In particular, the predicted population
can match the slope and scatter of the Tully-Fisher relation as well
as the properties of damped Ly$\alpha$ absorbers in QSO spectra.
We take their model as basis to derive the structural properties of
our N-body models of disk galaxies.
For definiteness, we briefly summarize the relevant assumptions and
equations.

\subsection{Dark haloes}

Using high-resolution N-body simulations,
\citet*[hereafter NFW]{NFW,NFW2} 
established 
that haloes formed by the 
gravitational clustering of cold dark matter 
exhibit a universal structure.
Suitably scaled, 
the density distribution of these dark matter haloes
does not depend on cosmology.
The NFW-profile is given by
\be
\rho(r)=\rho_{\rm crit.}\frac{\delta_{\rm c}}{ ( {r}/{r_{\rm s}})
\left(1+{r}/{r_{\rm s}} \right)^2},
\label{eqnfw}
\ee
where $\rho_{\rm crit.}$ is the background density at the time of the
halo formation, $r_{\rm s}$ is a scale
radius, and $\delta_{\rm c}$ is a characteristic overdensity.
Note that the slope of this profile is shallower than isothermal at
the center, and it gradually steepens outward to an asymptotic slope
of $-3$.
Following NFW, we define the {\it virial
radius} $r_{200}$ as the radius with mean overdensity 200, i.e.\ it
contains 
the {\em virial mass}
\be
M_{200}\equiv 200\rho_{\rm crit.}\,\frac{4\pi}{3}r_{200}^3\, ,
\ee
and we define the {\it concentration}
\be
c\equiv\frac{r_{200}}{r_{\rm c}}
\ee
of the halo. 
With these definitions, the characteristic overdensity is given by
\be
\delta_{\rm c}=\frac{200}{3}\,\frac{c^3}{\ln(1+c)-\frac{c}{1+c}}.
\ee
Further, let 
\be
v_{200}^2\equiv\frac{GM_{200}}{r_{200}}
\ee
be the circular velocity at the virial
radius.
Given the concentration $c$ and the Hubble constant $H(z)$,
the radial density profile of a halo may then be specified 
by anyone of the parameters
$v_{200}$, $r_{200}$, or $M_{200}$.
In particular, we have
\be
M_{200}=\frac{v_{200}^3}{10 G H(z) },  \;\;\;\; \mbox{and}\;\;\;\;\;
r_{200}=\frac{v_{200}}{10 H(z)} .
\ee

\subsection{Putting a disk into the halo}

We now put a stellar disk into an NFW halo according to the model of MMW.
This rests on four key assumptions:
\begin{enumerate}
\item
The mass $M_{\rm d}$ of the disk is a given fraction $m_{\rm d}$ of
the halo mass.
\item 
The spin $J_{\rm d}$ of the disk is a given fraction $j_{\rm d}$ of
the angular momentum $J$ of the halo.
\item
The disk has the structure of a thin exponential disk, and it 
is cold and centrifugally supported.
\item
Only disks that are dynamically stable against bar formation
correspond to observable disk galaxies.
\end{enumerate}

\noindent
The angular momentum $J$ of a halo with total energy $E$ 
is often characterized by the dimensionless
spin parameter
\be
\lambda=\frac{J |E|^{\frac{1}{2}}}{G M^{\frac{5}{2}}} .
\label{eqspin}
\ee
According to N-body results \citep{Wa92,Le98}, the distribution of
$\lambda$ is well approximated by
\be
p(\lambda)\,\dd \lambda =\frac{1}{(2\pi)^{1/2} \sigma}\exp\left[ -\frac{ (\ln\lambda
-\ln\ol{\lambda})^2}{2\sigma^2}\right] \frac{\dd \lambda}{\lambda}
\ee
with $\sigma=0.5$ and a typical value $\ol{\lambda}=0.05$. 
This distribution is practically independent of
cosmology, and of the mass and environment of the haloes \citep{Le98}.
The initial kinetic energy of the spherically symmetric halo 
may be computed by assuming
that all particles move 
around the center on circular orbits, with speed equal to the
circular velocity.
This `trick' results in 
\be
E_{\rm kin}=\frac{GM^2}{2r_{200}}f_{c},
\ee
where 
\be
f_{c}=\frac{c\left[ \frac{1}{2} -\frac{1}{2(1+c)^2}
-\frac{\ln(1+c)}{1+c}\right]}
{\left[ \ln(1+c) -\frac{c}{1+c}\right]^2} .
\ee
Using the virial relation $E=-E_{\rm kin}$, the angular momentum of
the halo then becomes
\be
J=\lambda G^{\frac{1}{2}} M^{\frac{3}{2}}
\left(\frac{2r_{200}}{f_{c}}\right)^{\frac{1}{2}} .
\label{eqspin2}
\ee

We now put a fraction $m_{\rm d}$ of the initial halo mass
into a thin stellar disk with an exponential surface density,
viz.
\be
\Sigma(R)=\Sigma_0 \exp\left(-\frac{R}{R_{\rm d}}\right)
\ee
with
$\Sigma_0={M_{\rm d}}/({2\pi R_{\rm d}^2})$.
Here $M_{\rm d}=m_{\rm d} M_{200}$ is the total mass of the disk and
$R_{\rm d}$ is
its 
scale radius.
The condition 
\be
J_{\rm d}=j_{\rm d} J
\label{eqxx1}
\ee 
will then determine the scale radius of the
disk, because
its spin is given by
\be
J_{\rm d}=M_{\rm d}\int_0^{r_{200}} \left(\frac{R}{R_{\rm d}}\right)^2
v_{\rm c}(R)\exp\left(-\frac{R}{R_{\rm d}}\right) \dd R ,
\ee
where 
the circular velocity $v_{\rm c}$ is the sum of two contributions,
namely
\be
v_{\rm c}^2(R)\equiv R\frac{\partial \Phi}{\partial R} = v_{\rm c,disk}^2(R) + v_{\rm c,dm}^2(R).
\label{eqxx3}
\ee

\subsection{Response of the dark matter profile}

We take the gas, that later forms the disk, to be initially
distributed just like the dark matter. However, the 
structure of the dark halo will be changed when the disk forms in its
center. We again follow MMW 
and assume that the dark matter reacts
adiabatically to the disk formation. In particular, we assume that the
spherical symmetry of the halo is retained, and that the angular momentum
of individual dark matter orbits is conserved.
This latter condition may be formulated as 
\be
r_{\rm i} M(r_{\rm i}) = r_{\rm f} M_{\rm f}(r_{\rm f}).
\ee
Here $r_{\rm i}$ and $r_{\rm f}$ are the initial and final radii 
of some dark matter mass shell, $M(r)$ gives the
initial NFW mass profile, and $M_{\rm f}(r)$ is the final cumulative
mass profile after the disk is formed. $M_{\rm f}(r)$ is the
sum of the cumulative mass of the disk and the dark mass inside the initial
radius, i.e.\ %
\be
M_{\rm f}(r_{\rm f}) =  M_{\rm d}(r_{\rm f})
+ (1-m_{\rm d}) M(r_{\rm i}) .
\label{eqxx2}
\ee
The 
final profile 
$M_{\rm h}(r)$
of the dark
matter halo is then given by
\be
M_{\rm h}(r)=M_{\rm f}(r)-M_{\rm d}(r).
\label{eqxx4}
\ee

For a given set of parameters $v_{200}$, $c$, $m_{\rm d}$, $j_{\rm
d}$, $\lambda$, and a formation redshift $z$, the above equations uniquely determine a disk
model. Note that in practice 
the scale length $R_{\rm d}$ of a disk needs to be
determined iteratively in order to satisfy equations (\ref{eqxx1}) and (\ref{eqxx2}).

\subsection{Including a bulge}

In many galaxies, including the Milky Way, a central bulge population
of stars is observed. For spirals like the Milky Way or of later type,
the bulge mass is less than 20\% of the disk mass. For this reason,
the dynamical importance of the bulge in these systems should be small.
However, there are also systems with a higher mass fraction in the
bulge. 
While most of the models in this study do not have a bulge, we still want to 
investigate its possible influence on our results.
Hence we here
generalize the above model to allow the option of a bulge.

Bulges appear to be flattened triaxial
systems, that may be partly supported by rotation.
However, \citet{He93d} found that it hardly matters for
the density and velocity structure of merger remnants
whether bulges are spinning or not.
For simplicity, we therefore 
neglect a possible flattening of the bulges and model them as
non-rotating spheroids with 
a spherical Hernquist profile of the form
\be
\rho_{\rm b}(r)=
\frac{M_{\rm b}}{2\pi}\frac{r_{\rm b}}{r(r_{\rm b}+r)^3} .
\ee

In analogy to the treatment of the disk, 
we assume that the bulge mass is a fraction
$m_{\rm b}$ of the halo mass. Since we take the bulge to be
non-rotating it has lost its specific angular momentum either to the
halo, or to the disk. 
We will assume that there is no angular
momentum transport between the disk and the dark halo, and none
between the disk and the bulge. In this case $j_{\rm d}=m_{\rm d}$.

For simplicity, we further assume that the bulge scale radius $r_{\rm b}$ is a
fraction $f_{\rm b}$ of that of the disk, i.e.\ $r_{\rm b}=f_{\rm
b} R_{\rm d} $. 
Note that the disk half mass radius is 
$1.678\, R_{\rm d}$, while that of the bulge is $2.414\, r_{\rm b}$.

It is then straight forward to generalize the above disk model to
accomodate the bulge. The circular velocity of equation (\ref{eqxx3}) gets an
additional contribution from the bulge, i.e.
\be
v_{\rm c}^2(R)=v_{\rm c,disk}^2(R) + v_{\rm c,dm}^2(R)  +  v_{\rm
c,b}^2(R) ,
\ee
with $v_{\rm c,b}^2(R)= G M_{\rm b}(R)/{R}$. Further, 
equation (\ref{eqxx2}) 
needs to be replaced by
\be
M_{\rm f}(r_{\rm f}) =  M_{\rm d}(r_{\rm f}) + M_{\rm b}(r_{\rm f})
+ (1-m_{\rm d}-m_{\rm b}) M(r_{\rm i}) ,
\ee
and the dark mass profile of equation (\ref{eqxx4}) now becomes
\be
M_{\rm h}(r)=M_{\rm f}(r)-M_{\rm d}(r)-M_{\rm b}(r) .
\ee

\section{N-body realizations}

\subsection{N-body realizations of model galaxies}

In order to construct near-equilibrium N-body realizations 
of our disk models, we need to initialize both positions and
velocities of particles according 
to the solution of the collisionless Boltzmann equation
(CBE)\footnote{also known as Vlasov equation.}.
While the first can easily be done according 
to the 
derived mass distributions for halo, bulge, and disk, the latter
is considerably more complicated.

Instead of attempting to solve the CBE directly, we
follow \citet{He93} and 
assume that the velocity distribution at a given point in space can be
sufficiently well approximated by a multivariate Gaussian. In this
case, only the first two moments of the velocity distribution are
needed. They can be obtained by taking moments of the CBE, a process
that leads
to a hierarchy of generalized
Jeans equations \citep{Ma94}.

For a static, axisymmetric system, the energy $E$ 
and the 
angular momentum component $L_z$ are conserved
along orbits. With the assumption that the distribution function
depends only on $E$ and $L_z$
one can show \citep{Ma94} that
the first
velocity moments are given by 
\be
\ol{v_R}=\ol{v_z}=\ol{v_R v_z}=\ol{v_z v_\phi}=\ol{v_R v_\phi}=0 ,
\ee
\be
\ol{v_R^2}=\ol{v_z^2}
\label{eqqq} ,
\ee
\be
\ol{v_z^2} = \frac{1}{\rho}\int_z^{\infty} \dd z' \, \rho(R,z')
\frac{\partial \Phi}{\partial z'}(R,z')
\label{momz} ,
\ee
\be
\ol{v^2_\phi}=\ol{v^2_R}+\frac{R}{\rho} \frac{\partial}{\partial 
R}(\rho\ol{v_R^2})
+v_{\rm c}^2
\label{equ5} ,
\ee
where the {\it azimuthal circular velocity} is defined as
\be
v_{\rm c}^2 \equiv R \frac{\partial \Phi}{\partial R}.
\ee
Not specified by the Jeans equations is the azimuthal streaming
$\ol{v_\phi}$, which can essentially be freely chosen in the context
of the above approximations. This reflects the fact that the distribution
function is even in $L_z$; the relative contribution of the parts with
positive and negative $L_z$ can be arbitrarily chosen.

We employ the assumption $f=f(E,L_z)$ for the dark matter halo and the
optional bulge. However, a realistic distribution function for the 
disk has a more complicated structure, and we will treat it slightly
differently, as described below.

\subsubsection{Structure of the disk}

Real stellar disks have a finite thickness. For their vertical 
structure we adopt the common choice
of Spitzer's isothermal sheet, viz.\ %
\be
\rho_{\rm d}(R,z) = \frac{\Sigma(R)}{2z_0} {\rm
sech}^2\left(\frac{z}{z_0}\right) .
\ee
Here the thickness $z_0$ of the disk sets its `temperature'. Most
spiral galaxies seem to be consistent with a constant vertical scale
length with a value of $z_0\simeq 0.2 R_{\rm d}$, which we will adopt in the following.

The distribution function of the disk depends on more than just two
conserved quantities, hence it is unrealistic to assume an isotropic
velocity dispersion. However, we will keep the assumption that the
velocity ellipsoid is aligned with the coordinate axes. Then 
equation (\ref{momz}) remains valid, 
and we use it to compute $\sigma_z^2(R,z)$.
Note that 
due to the radial variation of $\Sigma(R)$ and the presence of
the halo, the vertical velocity structure of the disk will not
be 
exactly
isothermal. 

We further employ the epicycle approximation \citep[chapter 4]{Bi87} to
relate the radial and azimuthal velocity dispersions by
\be
\sigma_\phi^2=\frac{\sigma_R^2}{\gamma^2}.
\ee
Here we have defined 
\be
\gamma^2\equiv \frac{4}{\kappa^2 R}\,\frac{\partial \Phi}{\partial R},
\ee
and the epicyclic frequency $\kappa$ as
\be
\kappa^2 \equiv \frac{3}{R} \,\frac{\partial \Phi}{\partial R} +
\frac{\partial^2 \Phi}{\partial R^2}.
\ee
The epicycle approximation also implies that the asymmetric drift is
small. We neglect it altogether and 
set the streaming velocity to be equal to the circular
velocity, i.e.\ $\ol{v_\phi} =  v_{\rm c}$.
For simplicity, we also continue to assume that 
$\sigma_R^2=\sigma_z^2$.
With these assumptions the velocity structure of the disk is fully specified.

Note that the thickness of the disk must be chosen large enough
to fulfil Toomre's stability criterion, which requires
\be
Q\equiv \frac{\sigma_R \kappa}{3.36 G \Sigma} > 1
\ee
to ensure local stability in differentially rotating disks.

For our models, the minimum value of $Q$ is about 1.4. Hence, we
could have taken somewhat colder disks, which might produce sharper
tidal tails.

\subsubsection{Rotation of the halo}

For consistency, we want to properly represent
the angular momentum carried by the dark matter, 
although we do not expect that halo rotation will have a strong 
influence on tail formation.

We model the streaming
velocity of the dark halo 
as some fixed fraction $f_{\rm s}$ of the local {\em azimuthal} circular
velocity, i.e.
\be
\ol{v_\phi}=f_{\rm s} v_{\rm c}.
\label{eq2}
\ee
If the specific angular momentum of the dark matter is conserved
during disk formation, the factor $f_{\rm s}$ stays fixed as well. Hence it
can be computed for the initial NFW-profile. For the streaming of
equation (\ref{eq2}) the initial angular momentum is
\be
J=\frac{2}{3}\, f_{\rm s} g_{c} \, \frac{G_{\,}^{\frac{1}{2}} M^{\frac{3}{2}}
r_{\rm s}^{\frac{1}{2}}}
{\left[\ln(1+c) -\frac{c}{1+c}\right]^{\frac{3}{2}}} , 
\ee
where $g_c$ is the integral
\be
g_c=\int_0^c \left[ \ln(1+x)-\frac{x}{1+x}\right]^{\frac{1}{2}}
\frac{x^{\frac{3}{2}}}{(1+x)^2} \dd x  .
\ee
Comparing this with equation (\ref{eqspin2}) we see that $f_{\rm s}$ is given by
\be
f_{\rm s}=\frac{3}{2}\lambda \left(\frac{2c}{f_{c}}\right)^{\frac{1}{2}}
g_c^{-1}
{\left[ \ln(1+c) -\frac{c}{1+c} \right]^{\frac{3}{2}}} .
\ee
The quantity $f_{\rm s}/\lambda$ varies only weakly, e.g.\ it takes
the value 4.5 for $c=1$, and reduces to
3.3 for $c=100$.

\subsubsection{Halo truncation}

For the halo we encounter a slight technical problem, since 
the cumulative mass distribution
of the NFW profile
actually diverges
for large $r$.
This is simply due to the fact that the NFW-profile 
in the form of equation (\ref{eqnfw}) is not valid 
out to arbitrarily large distances; it just provides a good fit 
to the profile up to about the virial radius. 

Instead of 
truncating the profile sharply at the virial radius 
we rather want to derive 
N-body models
where the density
fades out
smoothly.
For simplicity, we 
have chosen an exponential cut-off that sets in at
the virial radius and turns off the profile on
a scale $r_{\rm s}$, viz.
\be
\rho(r)=\frac{\rho_{\rm crit.}\delta_{\rm c}}{c(1+c)^2}
\left(\frac{r}{r_{200}}\right)^a 
\exp\left(-\frac{r-r_{200}}{r_{\rm s}}\right)
\ee
for $r>r_{200}$.
The power law exponent $a$ allows a smooth transition of the profile at
$r_{200}$. We select $a$ such that the logarithmic slope
\be
n=\left. r\frac{\dd}{\dd r}\ln \rho\right|_{r_{200}} =-\frac{1+3c}{1+c}
\ee
of the profile at the virial radius is continuous.
This implies
$a=c+n$.

Note that this truncation results in some additional halo mass beyond the
virial radius. The total mass $M_{\rm tot}$ is roughly 
10\% larger than $M_{200}$. However, we want to keep our definitions
of disk and bulge masses in terms of the virial mass. 
As a consequence 
we need to slightly modify equation (\ref{eqxx2}). It becomes
\[
M_{\rm f}(r_{\rm f}) =  M_{\rm d}(r_{\rm f}) + M_{\rm b}(r_{\rm f})
+ 
\left[1- \frac{M_{200}}{M_{\rm tot}}(m_{\rm d}+m_{\rm b})\right] M(r_{\rm i}) .
\nonumber
\]

\begin{figure*}
\bc
\resizebox{16cm}{!}{\includegraphics{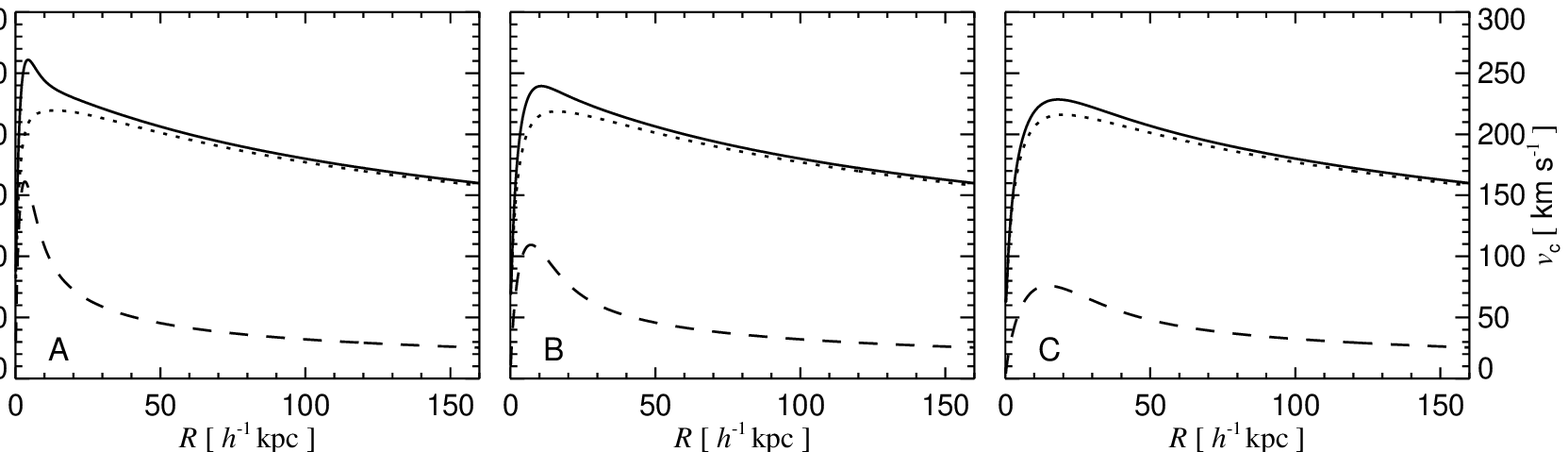}}\vspace*{0.5cm}\\
\resizebox{16cm}{!}{\includegraphics{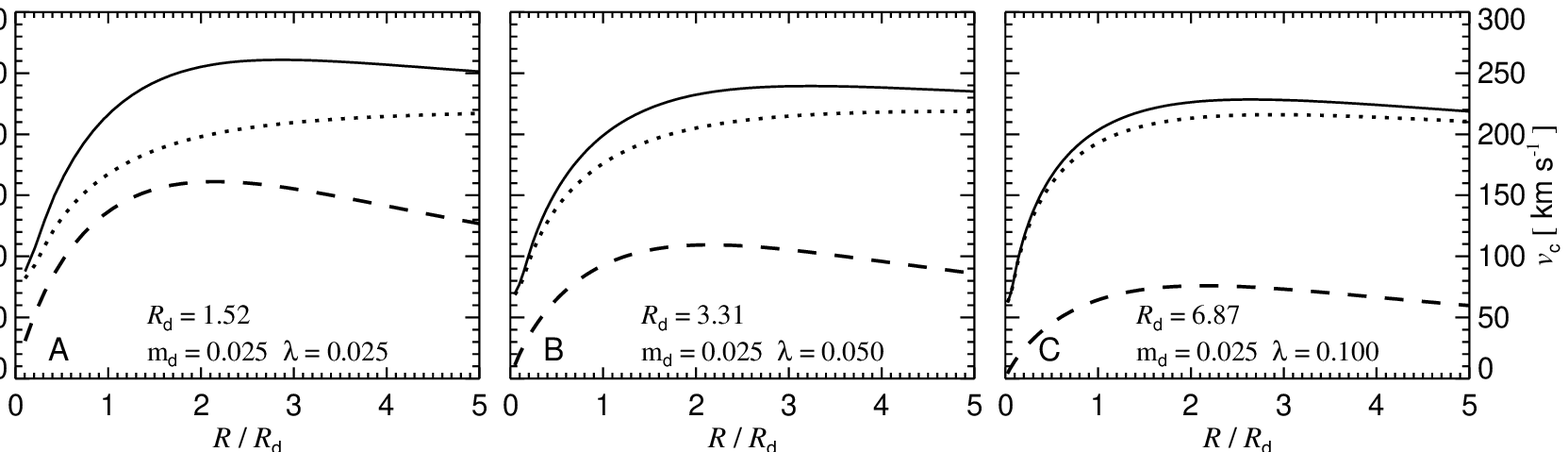}}\\
\resizebox{16cm}{!}{\includegraphics{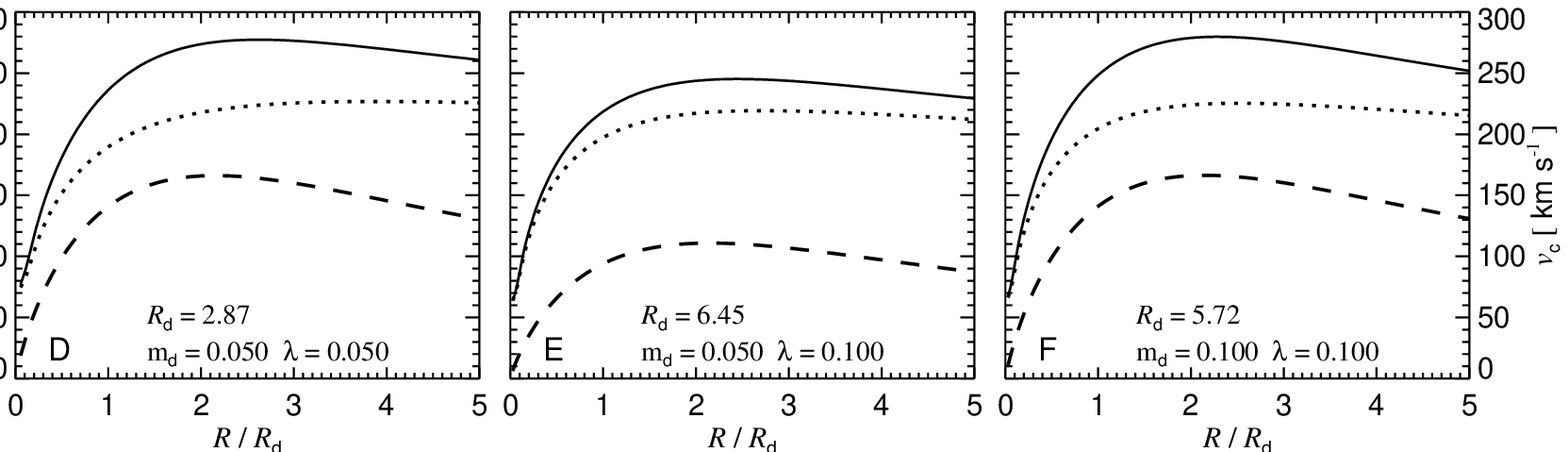}}\\
\caption{Rotation curves of disk models A to F. 
The top three panels 
show the full rotation curve of models A-C out to the virial radius. 
In the other six panels, the
radial coordinate is normalized to the disk scale
length, and we 
plot the inner rotation curve out to 5 disk scale lengths.
In each panel, the dotted curve gives the
contribution of the dark matter, the dashed that of the disk, and the
solid line is the total rotation curve.
In all the models, the total mass of the galaxies is equal,
and corresponds to an initial NFW dark matter profile with
$v_{200}=160\,\mbox{km s}^{-1}$ and $c=15$.
\label{rotcurves1}}
\ec
\end{figure*}

\subsubsection{Numerical procedure}

Finally, we briefly describe 
our computer code 
to set up a galaxy according to the above model.
The following steps are followed:

\begin{enumerate}
\item
Particle positions are initialized according to 
the density profiles  
for halo, disk, and bulge.
\item
We then compute the velocity dispersions on a fine logarithmic mesh 
in the $(R,z)$-plane.
For this purpose, we 
compute the integrals 
\be
\ol{v_z^2}=\frac{1}{\rho}\int_z^{\infty} \dd z'\, \rho \frac{\partial
\Phi}{\partial z'}
\label{eqint}
\ee
numerically at the grid points. 
This also determines $\ol{v_R^2}=\ol{v_z^2}$.
Note that the density refers only to the
component under consideration, while the potential is given by all the
material.
For the halo and the bulge,
we use equation (\ref{equ5})
and find $\ol{v_\phi^2}$ by
numerically differentiating $\rho \ol{v_R^2}$ in this
plane.  
For the disk, we use the epicycle approximation to determine
$\ol{\sigma_\phi^2}$, and we set the azimuthal streaming 
equal to the local circular velocity. The streaming of the halo is given
by equation (\ref{eq2}), while that of the bulge (if present)
is set to zero.

It should be noted that the integrals of equation (\ref{eqint}) 
require elaborate
numerical techniques, since the computation of the combined force field
is nontrivial.
\item
Finally, particle velocities are initialized by drawing random 
numbers from multivariate Gaussians with dispersions 
interpolated from the ($R,z$)-grid to the particle positions.
\end{enumerate}

This scheme is similar to that of \citet{He93}, 
although our numerical procedure
and treatment of the disk is somewhat different.

\section{Simulations}

\begin{table}  
\bc
\caption{Parameters of our basic set of six disk models. All of them
have the same total mass corresponding to $V_{200}=160\,\mbox{km}\,\mbox{s}^{-1}$,
and an initial halo concentration of $c=15$. From the possible
combinations
of $\lambda,m_{\rm d}\in\{0.025,0.5,0.1\}$ we consider only those
models that are stable according to the criterion
$\lambda\ge m_{\rm d}$. We assume that 
the disk material conserves its specific angular momentum,
i.e.\ $j_{\rm d}=m_{\rm d}$. 
The two tables illustrate our labeling of the models and the resulting
disk scale lengths.
Note that these models have no bulge.
\label{tab1}
}
\vspace{0.5cm}
\begin{tabular}{ccccc}
\hline
 & & & $\lambda$ & \\
 & & 0.025 & 0.05 & 0.1 \\
 & 0.025 & A & B & C \\
 $m_{\rm d}$ & 0.05 & & D & E \\
 & 0.1 & & & F\\
\hline
\end{tabular}
\vspace{0.8cm}
\begin{tabular}{ccccccc}
\hline
Model & A & B & C & D & E & F\\
$R_{\rm d}\;[\lu]$ & 
1.52 & 
3.31 & 
6.87 & 
2.87 & 
6.45 & 
5.72 \\
\hline
\end{tabular}
\ec
\end{table}

\begin{table} 
\bc
\caption{List of runs. The table gives the orbital angular momentum
of the different runs in terms of the minimum Keplerian separation
$R_{\rm kep}$. All the runs had an initial galaxy separation of 
$R_{\rm start}=320\,h^{-1}\mbox{kpc}$, and were set-up on a parabolic
encounter with zero total energy. Each of the runs is a collision
between identical disk models. The latter is 
specified by the initial character of the
labels. The bottom table gives the actual separation $R_{\rm min}$ 
of the disks in their first encounter.
\label{tab2}
}
\vspace{0.5cm}
\begin{tabular}{cccccc}
\hline
& \multicolumn{4}{c}{$R_{\rm kep}\;[\lu]$}  & \\
3.5 & 7.0 & 14.0 & 28.0 & 56.0 & 112.0 \\
\hline
A0 & A1 & A2 \\
B0 & B1 & B2 \\
C0 & C1 & C2 & C3 & C4 & C5\\
 &  & C2r \\
 &  & C2i \\
   & D1 &    \\
 &  & E2\\
 &  & F2\\
 & T1 &  \\
 & U1 &  \\
 & V1 &  \\
 & W1 &  \\
\hline
\end{tabular}
\vspace{0.3cm}
\begin{tabular}{cccccc}
\hline
\multicolumn{6}{c}{ $R_{\rm min}\;[\lu]$} \\
\hline
A0 & 6.4 & B2  & 23.7   & D1  & 10.8 \\
B0 & 8.3 & C2  & 22.4   & E2  & 18.2 \\
C0 & 7.4 & C2r & 19.4   & F2  & 22.4 \\
A1 & 10.5 & C2i & 21.2   & W1  & 10.9 \\
B1 & 14.0 & C3  & 37.4   & T1  & 8.3 \\
C1 & 13.6 & C4  & 58.2   & U1  & 9.4 \\
A2 & 21.3 & C5  & 109.6  & V1  & 10.8 \\
\hline
\end{tabular}

\ec
\end{table}

\begin{figure}
\bc
\resizebox{8cm}{!}
{\includegraphics{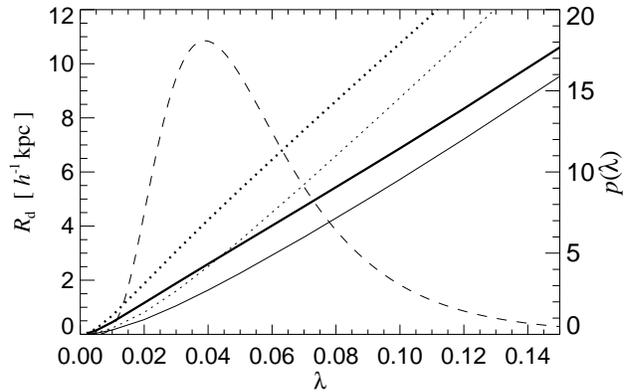}}\\
\caption{Sizes of disks.
The two solid lines show the scale length $R_{\rm d}$ of
disks as a function of the spin parameter $\lambda$. 
The solid lines are for haloes with $c=15$, 
while the dotted curves refer to $c=5$.
In both cases,
the heavy
curves are for $m_{\rm d}=0.025$, and the thin ones for $m_{\rm d}=0.1$.
The dashed line shows the distribution $p(\lambda)$ of $\lambda$ 
expected in CDM
cosmologies.
\label{rdlambda}}
\ec
\end{figure}

\subsection{Models}

\subsubsection{The basic disk models}

We have constructed a basic set of six disk models with a 
constant total mass corresponding to 
$V_{200}=160\,\mbox{km s}^{-1}$ and a concentration
of $c=15$. These models are those combinations of
$\lambda,m_{\rm d}\in\{0.025,0.5,0.1\}$ that result in stable 
cold disks, i.e.\ that have $\lambda\ge m_{\rm d}$ (the stability
criterion is discussed in more detail below). We label these
models A to F, as outlined in Table \ref{tab1}. Also given in this
Table are the resulting disk scale lengths. Note that we always assume
$j_{\rm d}=m_{\rm d}$, i.e.\ the specific angular momentum content of
the disk material is exactly conserved during disk
formation. In contrast to this, 
gas-dynamical simulations of disk formation \citep{Na94,Na97,We98}
have typically
led to a loss of angular momentum from the gas to the halo.
As a result, the disks formed in these simulations were much too small
to be identified with real spiral galaxies.
However, as all of the above authors note,
this angular momentum problem  may well 
be due to an insufficient treatment of feedback processes. 

Note that in the model of MMW, the structure of the disk galaxies
depends only on $\lambda' = (j_{\rm d}/m_{\rm d})\lambda$.
Hence, angular momentum loss from the disk ($j_{\rm d}<m_{\rm d}$)
has the same effect as lowering the value of $\lambda$.

\subsubsection{Rotation curves}

In Figure \ref{rotcurves1} 
we show the rotation curves of our six primary models
described in Table \ref{tab1}. For each model, 
we give 
the inner 
rotation curve out to 5 disk scale lengths, which is
about the accessible regime in most disk galaxies.
For models A to C, we also show
the full rotation
curve out to the virial radius of $160 \lu$.

Several interesting trends may be observed. In the models A, B, and C,
only the spin parameter is increased. This leads to larger disks with
roughly $R_{\rm d}\propto \lambda$. The dependence of $R_{\rm d}$ on
the spin parameter $\lambda$ is shown in Figure \ref{rdlambda}.
However, the smaller disks pull in
the dark matter more strongly, leading to a larger concentration of
the dark matter for smaller disks. This effect
reduces the differences between
the rotation curves when their radial coordinate is normalized to the
disk scale length.

On the other hand, for very small disk mass, the dark
matter profile will be nearly unaffected by the disk formation.
In this limit,
the disk stars behave more or less like test particles in the dark
matter potential, yet the size of the disk is still
determined by the halo spin parameter. Also, in this limiting case of a
massless disk it is quite clear that 
the mass ratio between halo and disk must be irrelevant for the
formation of stellar tidal tails.

\subsubsection{A model with a massive bulge}

\begin{figure}
\bc
\resizebox{8cm}{!}{\includegraphics{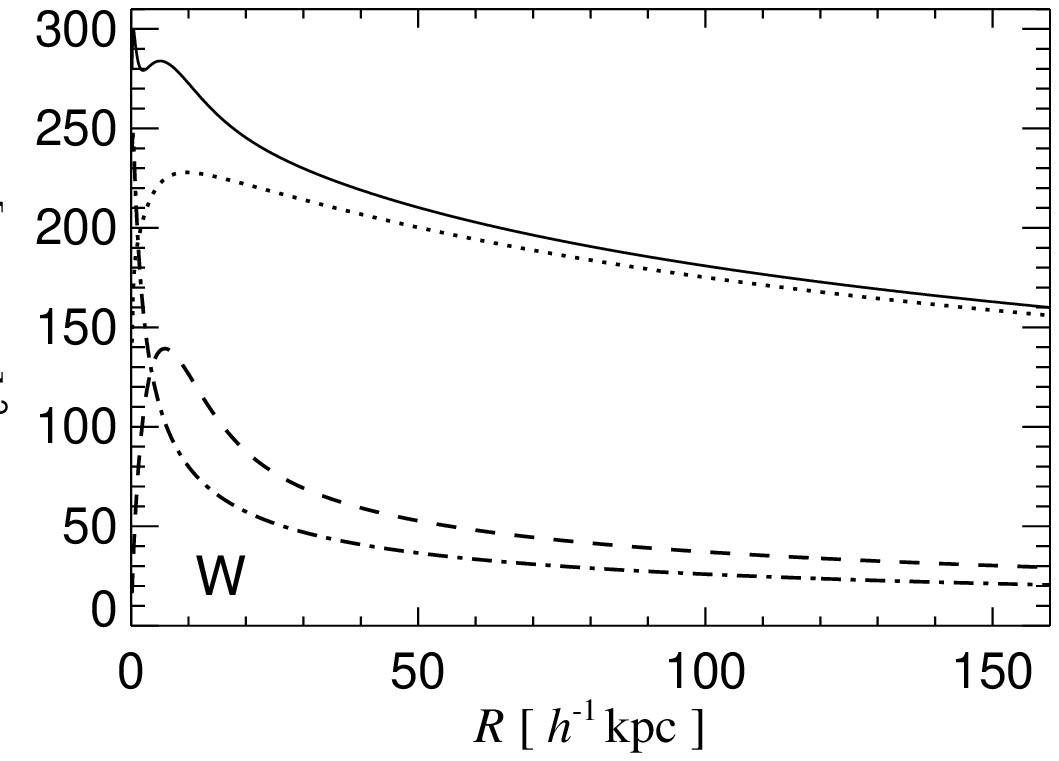}}\\
\resizebox{8cm}{!}{\includegraphics{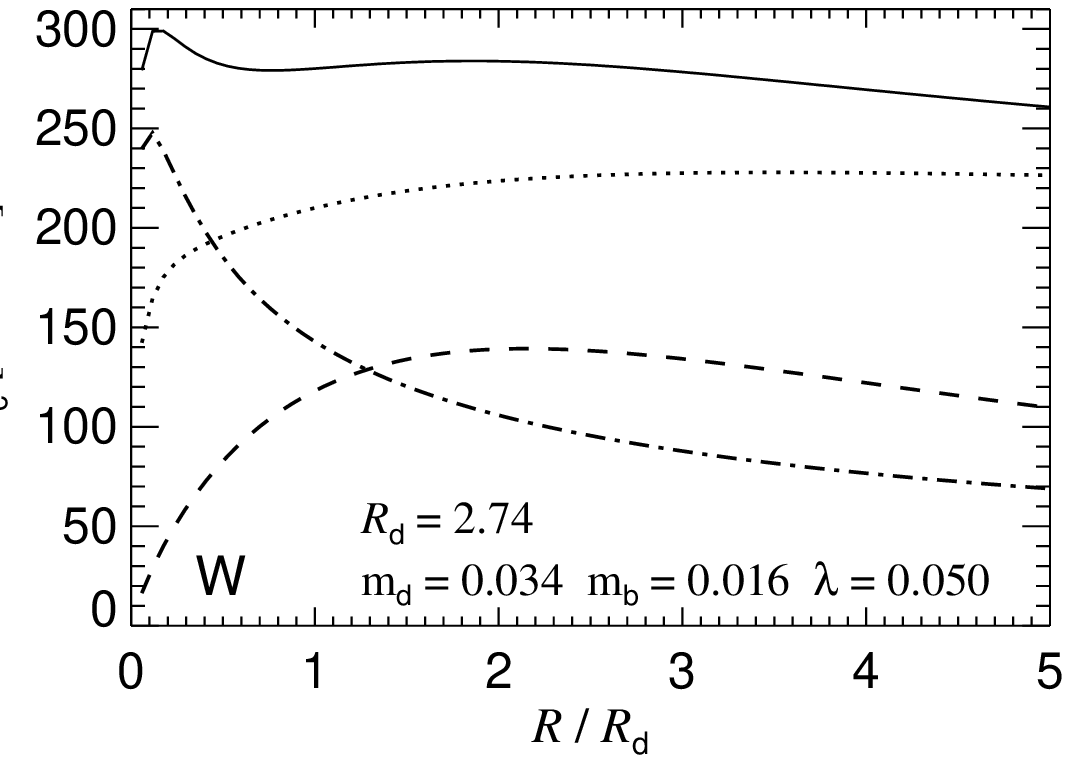}}\\
\caption{Rotation curves of the bulge model W.
The top panel shows 
the full rotation curve, and 
the lower panel displays the inner 
rotation curve out to 5 disk scale lengths. 
The dashed line is the contribution of the disk, the dot-dashed that
of the bulge, the dotted that of the dark halo, and the solid line
gives the total curve.
\label{rotcurvebulge}}
\ec
\end{figure}

We also consider a model with a massive bulge, designed
to have a similar rotation curve to
DMH's 
models,
and hence being more directly comparable to them than our standard models.
We also adopt their relatively high disk-to-bulge mass ratio of
2:1. In detail, our parameters for this model, which we call `W',
are $c=15$, $v_{200}=160\,\mbox{km s}^{-1}$,
$\lambda=0.05$, $m_{\rm d}=2/3\times 0.05$, $m_{\rm b}=1/3\times
0.05$, and $f_{\rm b}=0.1$. This results in the rotation curves shown
in Figure \ref{rotcurvebulge}.
While the inner rotation curve is very similar to that of
DMH,
the contribution of the dark matter is much more
important in our model, even at small radii. In particular, it is always much
larger than the contribution of the disk.

\subsubsection{The amount of dark mass in the disks}

\begin{figure}
\bc
\resizebox{8cm}{!}
{\includegraphics{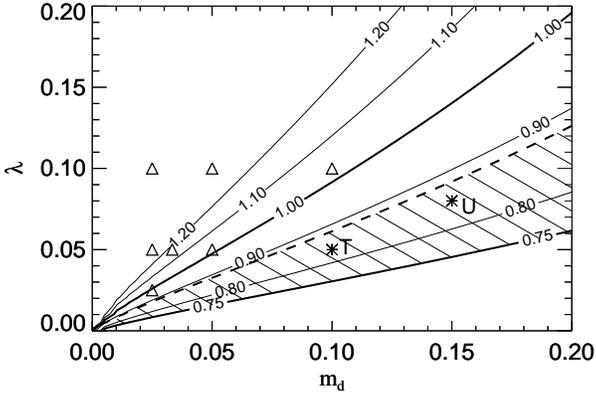}}\\
\caption{Contours of the disk stability parameter
$\epsilon_{\rm m}$ (solid lines) for $c=15$. According to Syer, Mao \& Mo
\protect\citet{Sy98} disks with $\epsilon_{\rm m}\ge 0.75$ should
be stable against bar formation, while the earlier work of Efstathiou,
Lake \& Negroponte \protect\citet{Ef82} 
gives the condition $\epsilon_{\rm m}\ge 1.1$.
Disk models lying below the thick dashed line are dominated by the disk
gravity at the 
maximum of the disk rotation curve, thus the hatched region shows the
parameter space that may contain stable disks which are not everywhere
dominated by dark matter. Two 
of our models, `T' and `U' (stars), lie in this region. The other
models are indicated as triangles.
\label{diskdom}}
\ec
\end{figure}

\begin{figure}
\bc
\resizebox{8cm}{!}
{\includegraphics{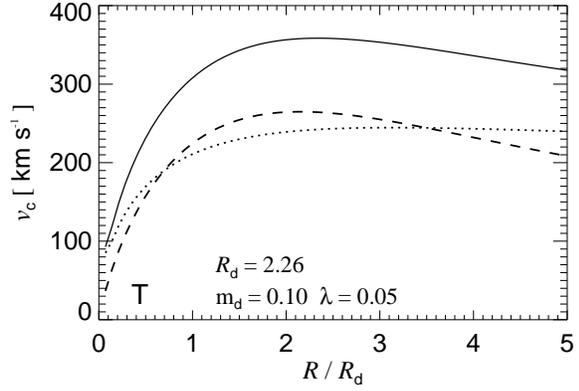}}\\
\resizebox{8cm}{!}
{\includegraphics{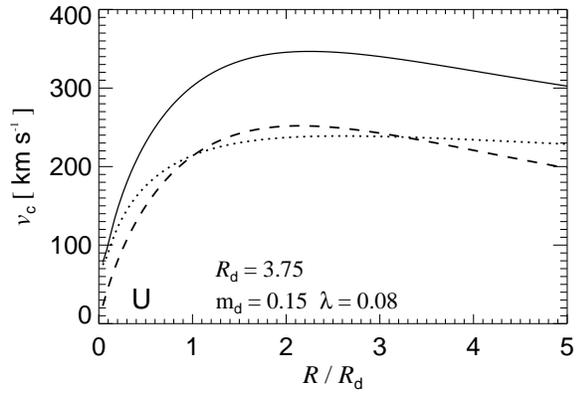}}\\
\resizebox{8cm}{!}
{\includegraphics{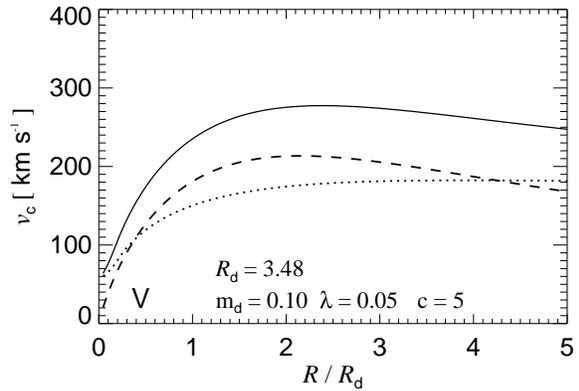}}\\
\caption{Inner rotation curves
of the models T, U, and V out to 5 disk scale lengths.
The dashed line is the contribution of the disk, the dotted 
that of the dark halo, and the solid line
gives the total rotation curve. 
\label{rotcurveT}}
\ec
\end{figure}

As the rotation curves in Fig.\ \ref{rotcurves1} show, all our `basic' disk 
models A-F are
gravitationally dominated by dark matter, even in the innermost
regions of the disks. While 
the presence of a dark matter halo 
has been convincingly demonstrated
by the flatness 
of observed rotation curves,
there is still a controversy about the amount of dark matter in the
inner regions of disk galaxies.

This controversy has arisen, 
because the decomposition of an observed rotation curve 
into a stellar and a dark matter
component is rather ambiguous, since the result depends
strongly on 
assumptions about the dark matter
profile and the mass-to-light ratio \citep{Na98}.
Traditionally, rotation curves have therefore 
been fitted using the `maximum-disk' hypothesis, 
i.e. the largest mass possible is assigned to the disk consistent
with the rotation curve.
The recent work of \citet{De98} 
suggests that
the central density of dark matter in barred galaxies should be low,
thus supporting the maximum-disk hypothesis.
However, others \citep[e.g.][]{Kr95} 
maintain
that the contribution of dark matter 
in the inner regions of disk
galaxies must be substantial.
There is also recent observational work 
that supports this assertion \citep{Qu98}. 
We also note that 
even if the inner rotation curve of a galaxy 
can be well accounted for by a
disk component alone, this does not provide evidence for the absence
of dark matter in the inner parts of the disk.

The theoretical results employed in this work 
predict dark matter profiles with a central density cusp, 
and a resulting strong
contribution of dark matter in the inner disks. 
We now briefly discuss, to what 
extent the model of MMW 
may also accommodate
galaxies that are maximum-disk, or at least somewhat closer to it.
In order to make the self-gravity of the disk more important, we can
either reduce $\lambda$ (making the disk smaller), increase $m_{\rm d}$
(making the disk heavier), or lower $c$ (reducing the concentration of
the halo). 
However, for fixed $m_{\rm d}$ and fixed $c$, disk-stability 
poses a lower limit on $\lambda$. 
Thin, fully self-gravitating
disks have been known to be violently bar-unstable for a long time, a
fact that suggested to \citet{Os73} that there
must be dark matter that stabilizes the disks. Later, \citet*{Ef82} 
used N-body simulations 
to derive the stability criterion 
$\epsilon_{\rm m} \ge 1.1$
for the 
disk, where
\be
\epsilon_{\rm m}\equiv
\frac{v_{\rm max}}{ (G M_{\rm d}/R_{\rm d})^{1/2}}, 
\ee
and
$v_{\rm max}$ is the maximum rotation velocity.
Recently, \citet*{Sy98}
confirmed that $\epsilon_{\rm m}$ is a good diagnostic for
bar-instability, although they found a somewhat weaker 
stability criterion,
$\epsilon_{\rm m} \ge 0.75$. 

In Figure \ref{diskdom} we show contours of $\epsilon_{\rm m}$
in the $m_{\rm d}$-$\lambda$ plane. Also shown is the region, where 
the disk gravity 
at the
maximum of the disk rotation curve
is larger than the contribution by the dark matter. 
There is thus a small region of parameter space (hatched) where 
the galaxies are
disk-dominated, but where they
should be still stable against bar formation according to
\citet{Sy98}. 
Incidentally, $\epsilon_{\rm m}\ge 1.0$ corresponds
very closely to the condition $\lambda\ge m_{\rm d}$, the choice
we employed so far.

Note that lowering $\lambda$ may not only render a disk unstable, 
it will also
make it substantially smaller. 
Hence, the observed sizes of disk galaxies can also provide 
a lower bound on $\lambda$. Indeed, the results of MMW
suggest that disks become too small if they lose a substantial
part of their angular momentum to the dark halo. Since
these arguments disfavour low $\lambda$, one may rather try to increase
$m_{\rm d}$ to make the gravity of the disk more important.
In principle, we
expect that the universal cosmic baryon fraction poses an upper limit
on $m_{\rm d}$,
while the actual value of $m_{\rm d}$ could be a lot smaller if the
efficiency of disk formation is low.
Taking a big bang nucleosynthesis value of
$\Omega_{\rm B}\simeq 0.015 h^{-2}$ \citep{Co95} for the baryon density, $m_{\rm
d}$ should be smaller than 0.06 in a critical density universe with 
a Hubble constant of
$h=0.5$. However,
clusters of galaxies suggest that the baryon fraction is larger by at
least a factor of three \citep{Wh93}. Note that in a 
low density universe this
can be reconciled with cosmic nucleosynthesis. If $\Omega_0$ is as low as
0.2, the limit on $m_{\rm d}$ goes up to about 0.15-0.2.

We examine these possibilities to a limited extent
with three additional models
which are disk-dominated 
in the inner regions. It is interesting to see 
how their tidal tails fit
into the systematic properties of the other models.
We label one of these models `T', and give it the parameters
$m_{\rm d}=0.1$, $\lambda=0.05$, $c=15$, $v_{200}=160\,\mbox{km
s}^{-1}$, and $m_{\rm b}=0$.
For a further model, called `U', we instead adopt $m_{\rm d}=0.15$,
$\lambda=0.08$, i.e.\ 
here we make the disk substantially more massive. Finally, we consider
a model `V' with a smaller concentration of the halo. Here we use 
$c=5$, $m_{\rm d}=0.1$, and $\lambda=0.05$. Hence this model 
is consistent with the value of
$c\simeq 5$ favoured by \citet{Na98} in
a
recent analysis of a sample of spiral galaxies.
Note that such 
low concentrations are theoretically expected for flat, 
low-density universes.

The inner rotation curves of these models are shown in Fig.\ \ref{rotcurveT}. 
The stability parameter for them is 
$\epsilon_{\rm m}\simeq 0.84$.
Hence they lie in the hatched region of
Figure \ref{diskdom}.
In contrast to the other models, we here chose
$\sigma_{R}=2.0\,\sigma_z$ for the velocity structure of the disk to
prevent it developing a bar before the galaxies collide. This raises
$Q$ to about 2.0, while it would have been $Q\simeq1.0$ for our conventional choice
for $\sigma_{R}$.

\subsection{Collision simulations}

The most favourable condition for making tidal tails are prograde
encounters where the spin vectors of the disks are aligned with the
orbital angular momentum. In this situation, the approximate resonance
between the disk rotation and the orbital angular frequency amplifies
the perturbation of particle orbits on the far sides of the disks,
since they stay for a longer time in the region of the strongest tidal
field.

Since we here try to achieve as prominent tails as possible, we
usually set up
our disk-disk collisions on prograde parabolic orbits. For simplicity
we run only symmetric encounters between pairs of identical models; that is
we collide model A with A, B with B, and so on.

We always chose the initial separation of the galaxies to be twice 
the virial
radius. i.e.\ $R_{\rm start}=320\lu$. 
The remaining undetermined orbital parameter is the 
orbital angular momentum.
We specify it in terms of the minimum separation $R_{\rm
kep}$ the 
galaxies would reach if they were point masses moving on the 
corresponding Keplerian orbit.
In reality, once the galaxies overlap they will start to deviate from
this trajectory due to dynamical friction. As a consequence, the actual
separation $R_{\rm min}$ of the galaxies in their first encounter will
generally be larger than $R_{\rm kep}$.

We examined three main choices for $R_{\rm kep}$, $3.5$, $7$ and
$14\lu$. For each of the models A, B, and C, we have run all three of these
combinations, 
while
we restricted ourselves to just one `impact parameter' for the other
models. Additionally,
we simulated a set of wider encounters with $R_{\rm kep}=28$, 56,
112$\lu$ 
for the C model.
Runs labeled `A0', `B0', etc.\ refer to $R_{\rm kep}=3.5\lu$,
those containing the digits 1 or 2 to $7\lu$ and $14\lu$,
respectively. 

We also simulated two additional versions of run C2 where the disks
do not 
have a prograde orientation. 
In the collision `C2r' both disks are retrograde,
i.e. their spins are just flipped, while in the model `C2i' they
are inclined by $90^{\rm o}$ relative to the orbital plane.

With respect to the bulge model and the disk-dominated models, 
we have only run one simulation in each case (`W1', `T1', `U1', `V1'), 
an encounter at $R_{\rm kep}=7\lu$.
Table \ref{tab2} gives an overview of all 
these runs. Also shown in this table are the actual minimum 
separations $R_{\rm min}$ of the centers of the disks in
their first encounter. 
We here 
defined the center of a disk as its densest point, and
used a kernel interpolation like in smoothed particle hydrodynamics (SPH)
to estimate the density of particles.
However, using simply the
center-of-mass of the indiviual disks gives similar results.

When the galaxies start to overlap, the interaction potential between
them becomes
shallower than that of the corresponding point masses. 
This effect will make the orbits wider than the Keplerian
expectation. However, the galaxies are also slowed down by 
dynamical friction, an effect that brings 
the galaxies closer together. 
Both mechanisms compete with each other, and their relative 
strength depends on
the distribution of mass inside the galaxy. As Table \ref{tab2}
shows, the minimum 
separations $R_{\rm min}$ are usually somewhat larger than the
corresponding Keplerian value $R_{\rm kep}$.
Note however, that the measurement of $R_{\rm min}$ has an uncertainty of
order $\pm 1\lu$,
because we stored only a limited number of
output times.

\subsection{Numerical techniques}

All the simulations in this work have been run with our  
\gadget-code ({\bf GA}laxies with {\bf D}ark matter and {\bf G}as
int{\bf E}rac{\bf T}).  It is a 
newly written SPH-code in C, specifically designed 
for the simulation of galaxy
formation and interaction problems.
The gravitational interaction is either computed with the
special-purpose hardware \grape\ (if available) or with a \tree\ code.
Time integration is performed with a multi-timelevel leapfrog
integrator, with 
the thermal energy equation being integrated
semi-implicitly.
Time-critical routines (i.e.\ the force computation and neighbour
search with the \tree\ or \grape) 
have been profiled and optimized extensively. 

In this work, the SPH part of  \gadget\ is not used; we  just treat dark
matter and stellar material as 
collisionless particles.
Futher details of \gadget\ and an application of its SPH-capabilities will be
described in future work.

For all of the basic models A to F, and for T to V,
we used 20000 particles to represent each disk, and
30000 particles for each halo, hence each simulation had a total of
$100000$ particles. We chose a gravitational softening length of
$0.4\lu$ for the dark matter, and $0.1\lu$ for the disk. 
Time integration was performed with high enough accuracy, such that the
total energy was conserved to better than $0.8\%$ in all runs.

For the bulge model W, we used an additional 10000 particles for each
bulge, and we employed a softening of $0.1\lu$ for the bulge as well.

Some of the models have been integrated using \grape\ (A0, B0, B2, E2,
F2, T1, U1, W1), 
the others with the \tree\ code. For the 
latter we used the cell opening criterion of \citet{Du96b} with
$\theta=1.0$, we included quadrupole moments, and we matched the spline
softening of the \tree\ code to the Plummer softening of  \grape\ cited above.

Each simulation was run for 2.6 internal time units, or 0.26 Hubble
times, corresponding to $2.54\times 10^9\,h^{-1}\mbox{yr}$.
At this point of time, the merger remnants are not yet fully relaxed,
but the tidal tails have already largely decayed.

\section{Results}

\subsection{Dynamical evolution of the models}

\begin{figure*}
\bc
{\includegraphics{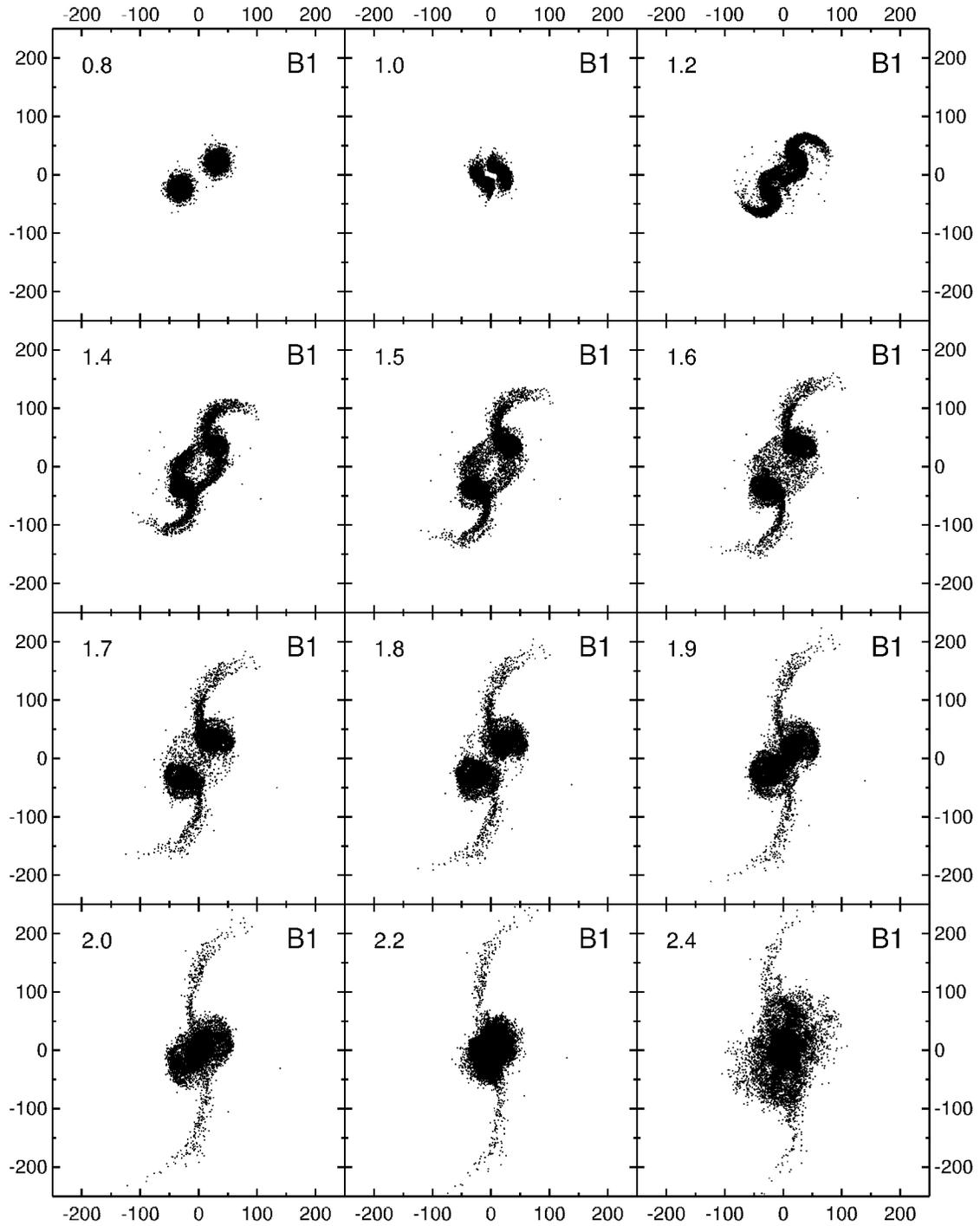}}\\
\caption{Time evolution of run B1. 
The panels show the disk particles projected onto the orbital
plane. The length units labeling the axes are given in $\lu$. 
The elapsed time since the start of the simulation (upper left
corners)
is given in units of 0.1 Hubble times, or $9.8\times 10^8\,h^{-1}{\rm yr}$.
\label{evolvA}}
\ec
\end{figure*}

Figure \ref{evolvA}  
shows a representative 
example of the time
evolution of one of our runs (B1).
Overall, the models follow the well-known behaviour of close
encounters of pairs of disk galaxies. 
When the galaxies reach
orbital centre, violent tidal forces induce a bar
instability that quickly 
transforms the disks into a pair of open bisymmetric spirals.  
Simultaneously, disk material from the far side of the encounter is
ejected by the tidal field into arcing trajectories that later form 
tidal arms. Material from the near side is drawn towards the
companion, giving rise to bridges between the galaxies as
they temporarily separate again. While the bridges are destroyed
when the galaxies come back together for a second time, the tails can
survive and grow for a longer time in the relatively quiet regions
of the outer potential.

Nevertheless, the dynamical evolution of the tidal tails is quite
rapid. 
After their
initial phase of expansion, the most strongly bound 
material in the inner region of the tail 
quickly starts to rain back onto the merging
pair. Eventually, this also happens to 
material progressively further out, such that
the surface density and prominence of the tidal tails quickly decrease
with time.

\begin{figure*}
\bc
\noindent{\includegraphics{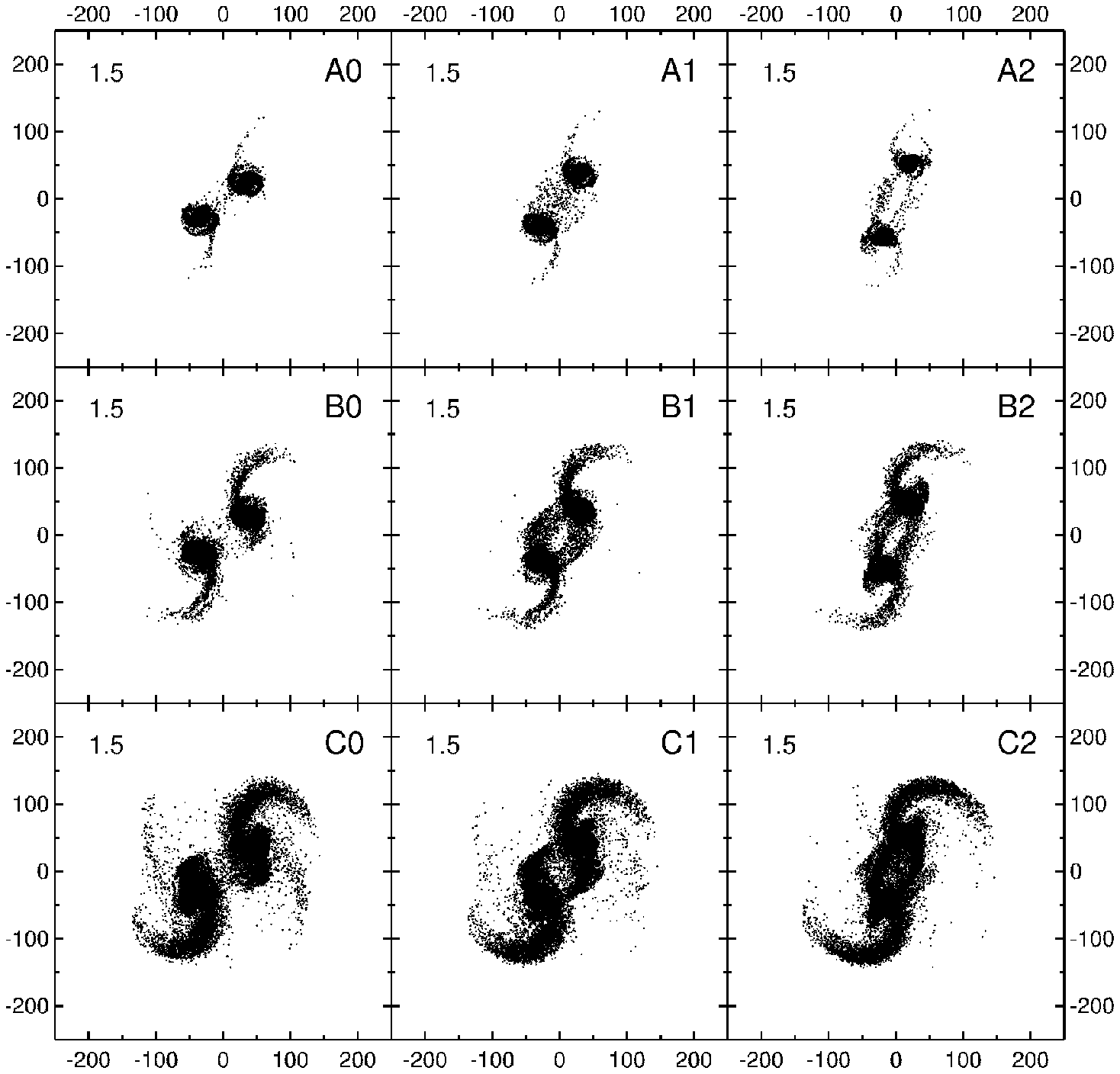}}\vspace*{-0.4cm}\\
\noindent{\includegraphics{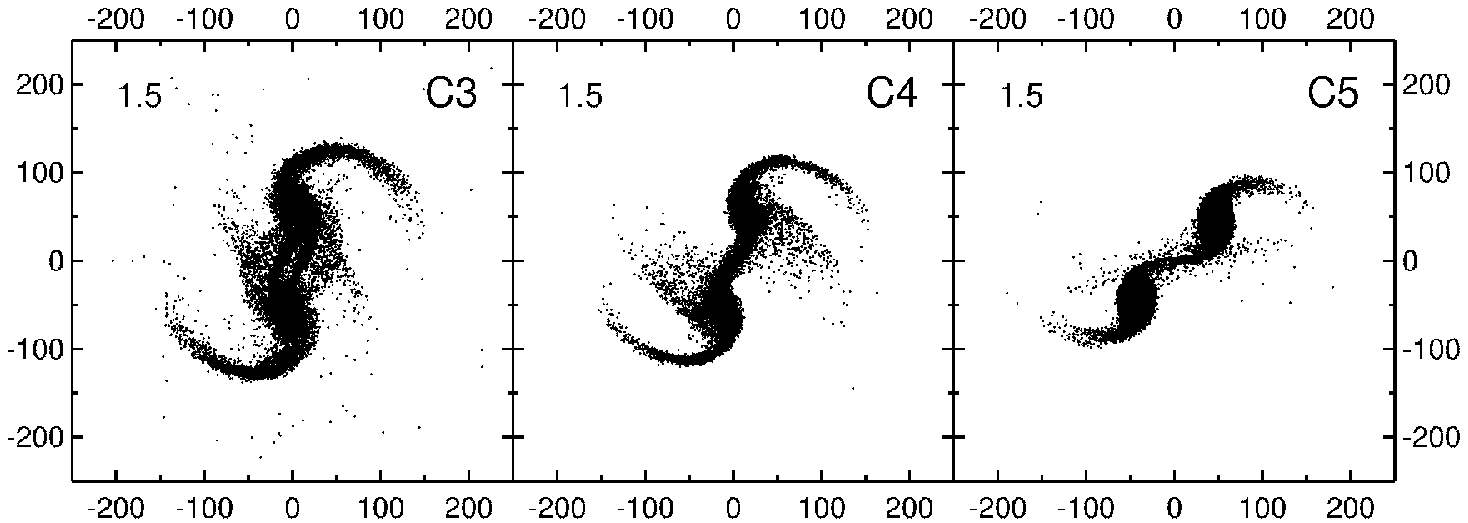}}\\
\caption{Comparison of tidal tails between different runs. All the
panels show the disk particles projected onto the orbital plane at the
same time since the start of the simulation.
The top 9 panels 
display runs that involve the disk models A-C for the set of 
impact parameters $R_{\rm kep}=3.5$, 7, and 14$\lu$ (indicated by the  
digits 0, 1, and 2, respectively, in the labels of the runs).
From A to C, the spin
parameter $\lambda$ increases in the sequence 0.025, 0.05, and 0.1,
but all three models have an equal disk mass given by $m_{\rm
d}=0.025$.
In the lower three panels we show additional collisons of model C with
wider impact parameters in the sequence $R_{\rm kep}=28$, 56, and 112$\lu$.
\label{tailcomp1}}
\ec
\end{figure*}

\begin{figure*}
\bc
\noindent{\includegraphics{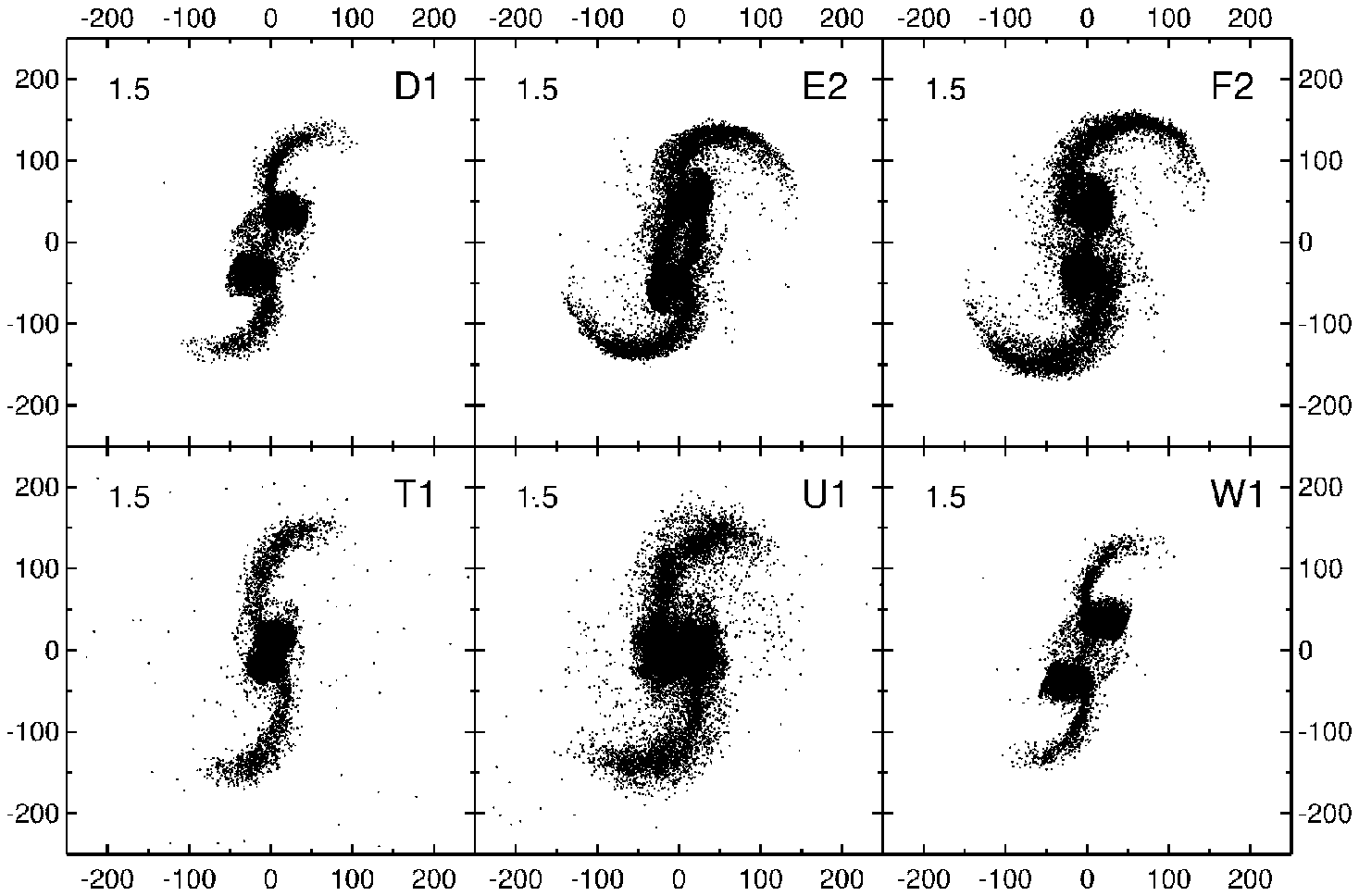}}\\
\caption{Comparison of tidal tails between different runs. 
The panels show the disk particles projected onto the orbital plane at the
same time since the start of the simulation.
Runs D1 and E2 involve models with twice as heavy disks ($m_{\rm
d}=0.05$) than A-C, while F2 has $m_{\rm d}=0.1$. 
The D-model has a spin parameter of
$\lambda=0.05$, while the E- and F-disks have $\lambda=0.1$.
The T ($m_{\rm d}=0.1$, $\lambda=0.05$) and U ($m_{\rm d}=0.15$, $\lambda=0.08$)
models are disk-dominated in their inner regions, while model W
($m_{\rm d}=0.034$, $m_{\rm b}=0.016$ 
$\lambda=0.05$) also
contains a bulge.
The digit in the label of the runs parameterizes the impact
parameter; 0, 1, and 2 are for $R_{\rm kep}=3.5$, 7, or 14$\lu$, respectively.
\label{tailcomp2}}
\ec
\end{figure*}

\subsection{Comparison of tidal tails}

Depending on the strength of the tidal response, 
the tidal tails can
contain a varying amount of mass, and reach different lengths. 
By comparing the time evolution of the models A to C, such 
differences are readily
apparent. For example, 
the tails of the larger disks of run C1 are much more massive and
prominent than those of run B1, while the tails of the small disks 
of simulation A1 are rather thin and anemic. However, the spatial extent of
the tails is quite comparable in the models. 
When normalized to the initial disk scale
length the anemic tails of the A-models are even longer than those of
the C-models.

These trends are clearly visible in Figures \ref{tailcomp1} and  
\ref{tailcomp2}, 
where we compare
different runs at the same time, approximately corresponding to the
moment when the tails are most impressive.
Note however, that due to the rapid evolution of 
the morphology of the tidal tails
it is not easy to compare different models at exactly equivalent
times of dynamical evolution.

When models with different impact parameters are compared, some
finer trends in the tail morphology 
may be observed. With growing impact parameter (labels $0\to 2$),
the bridges between the galaxies become more pronounced, and the tails
are slightly more curved. This is related to the larger orbital
angular momentum of these encounters.

Note that the runs A, B, and C of Figure \ref{tailcomp1} all have a large
halo-to-disk mass ratio of more than 40:1, i.e.\ according to 
DMH none of these models should have produced
prominent tails.
Nevertheless, 
the spatial extent of the tails is often quite
large. For example, the B-models with a disk scale length of $R_{\rm
d}\approx 3.5 \lu$ produce tails reaching $200\lu$ in length, i.e.\
about 60 times the original disk scale length.

When tails of models with different halo-to-disk mass
ratio but equal spin parameter are compared, it becomes clear that the mass ratio
cannot be the relevant parameter that decides whether tails 
form or not. For example, the tails of runs E2 and F2 in Figure 
\ref{tailcomp1} may be compared to the ones of run C2. 
Despite a variation of the mass ratio by a
factor 4 or so, the tails are almost equally strong in these three
simulations.
This is clearly due to the approximately equal size of the disks in 
these models. Because the dark matter is gravitationally
dominant even in the regions of the disks, the disk stars behave
almost like test particles in the gravitational potential of the dark
halo. In this limiting case it is clear, that 
only the {\em location} of the disk material inside the dark halo
determines
the disk response, i.e.\ it 
is the relative size of disk and dark halo that matters.

\subsection{An indicator for tidal response}

As we have seen above, knowledge of the
halo-to-disk mass ratio is by no means sufficient to predict how
prone a particular galaxy model is to tail formation. MMW suggested 
using the quantity
\be
{\cal E}=\left[\frac{v_{\rm e}(R)}{v_{\rm c}(R)}\right]^2
\ee
as a more suitable indicator. ${\cal E}$ compares the depth of the
potential well with the specific kinetic energy of the disk
material. The quantity ${\cal E}$ also arises, when one tries to estimate
the relative increase $\Delta E$ of specific kinetic
energy of disk stars in a nearly head-on encounter of identical disk
models. 
In this situation one finds (Binney \& Tremaine 1987; Mo, Mao \& White 1998)
\be
\Delta E /v_{\rm c}^2 \simeq 
\left(\frac{v_{\rm c}}{v_{\rm e}}\right)^2 = 1/{\cal E}.
\ee
This suggests the use of ${\cal E}$ as an indicator for the
susceptibility of a disk model
to tidal perturbations. In order to obtain a typical value for 
${\cal E}$ we evaluate it at $R=2R_{\rm d}$, which is about the half
mass radius of the disk.

\begin{figure}
\bc
\resizebox{8cm}{!}{\includegraphics{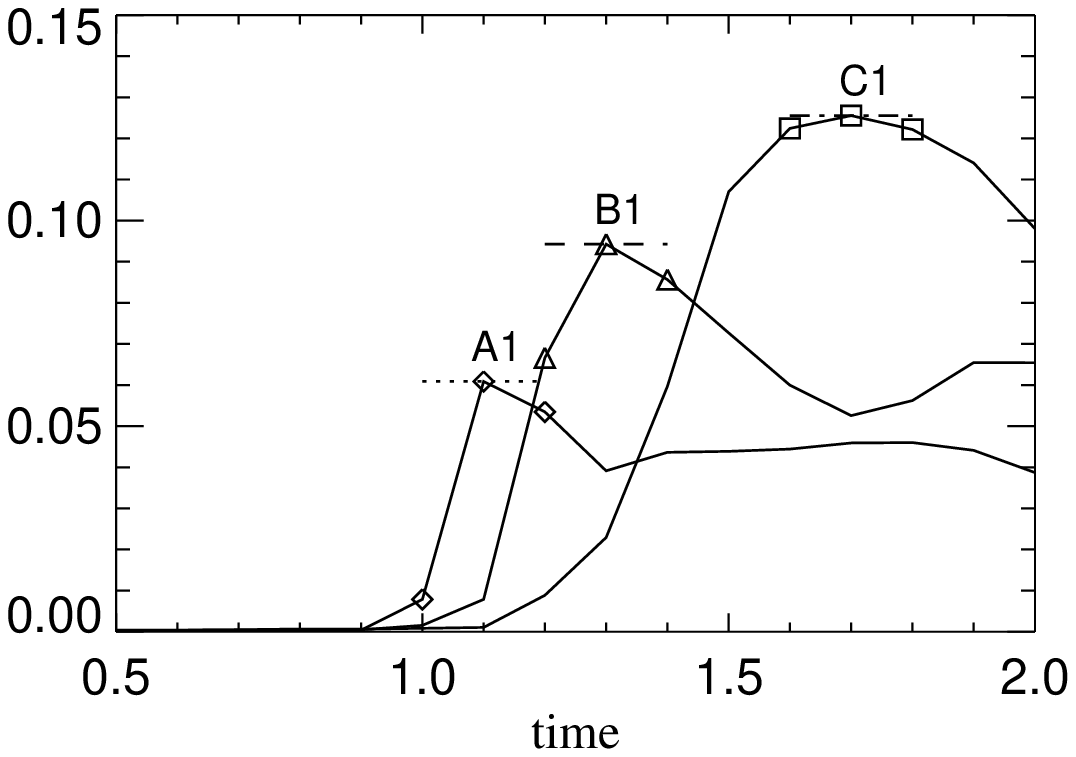}}\\
\caption{Tidal response $T$ as a function of time for three selected
runs. 
We define $T$ as the mass fraction of each disk that reaches a
distance of more than $10R_{\rm d}$ to its center-of-mass.
Also indicated as horizontal lines are the values of $T_{\rm eff}$
that we take as measure for the tidal response of the disk.
Since we only stored simulation outputs with a spacing of  
$0.1$ time units, the measurement of $T_{\rm eff}$
is slightly uncertain.
\label{resexamp}}
\ec
\end{figure}

\begin{figure*}
\bc
{\includegraphics{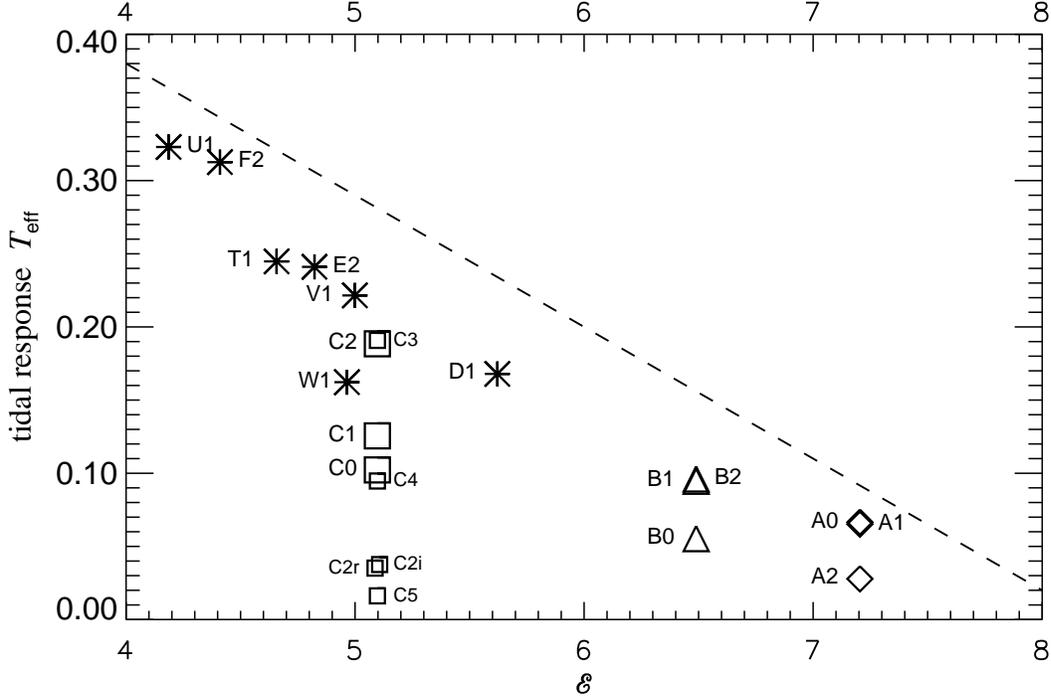}}\\
\caption{Tidal response $T_{\rm eff}$ 
of the different runs versus the value 
of ${\cal E}$ for the corresponding disk model, 
where ${\cal E}$ is evaluated at $R=2R_{\rm d}$.
The measurement of
$T_{\rm eff}$ is slightly uncertain, and 
$T_{\rm eff}$ also depends on details of the orbital parameters of
the encounter. However, our runs are designed to produce very nearly 
the strongest tails possible for a given disk model. 
At fixed ${\cal E}$, it should therefore 
be hard to find a model that gives a higher value 
for $T_{\rm eff}$ than the maximum of our runs. In other words,
there should be no model in the region above the indicated dashed line.
\label{response}}
\ec
\end{figure*}

\begin{figure*}
\bc
\resizebox{8cm}{!}
{\includegraphics{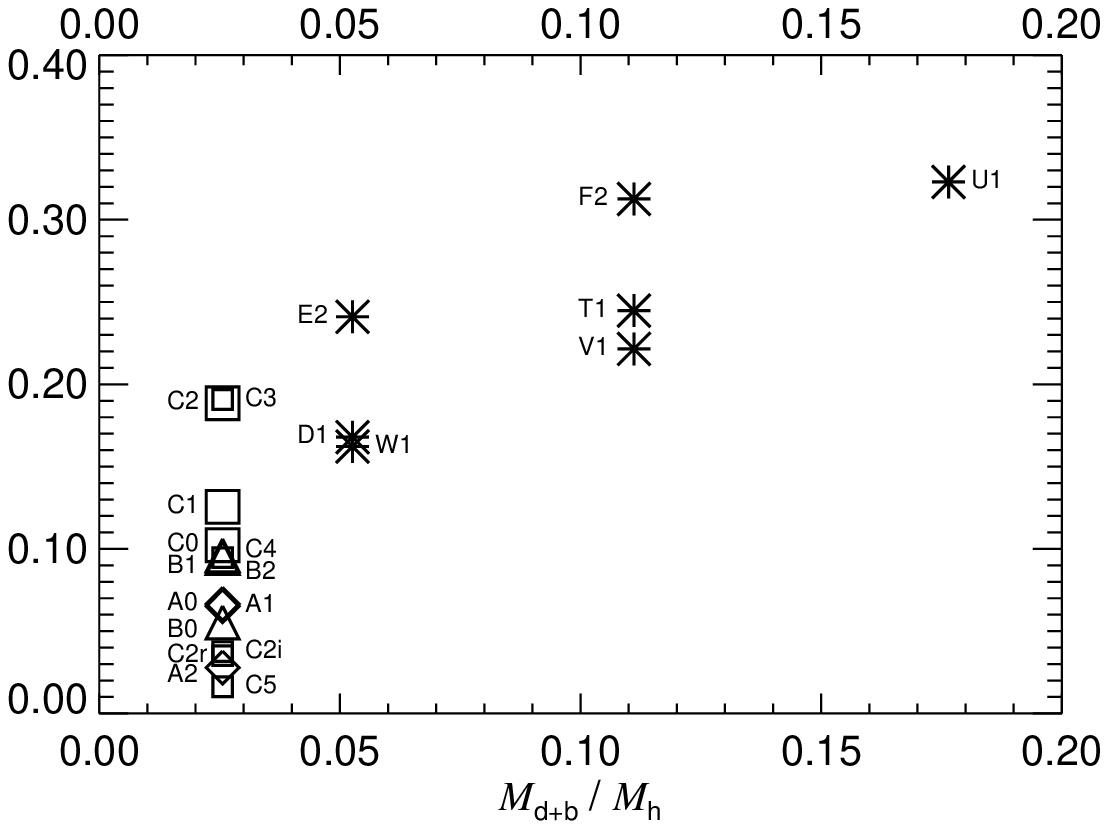}}
\resizebox{8cm}{!}
{\includegraphics{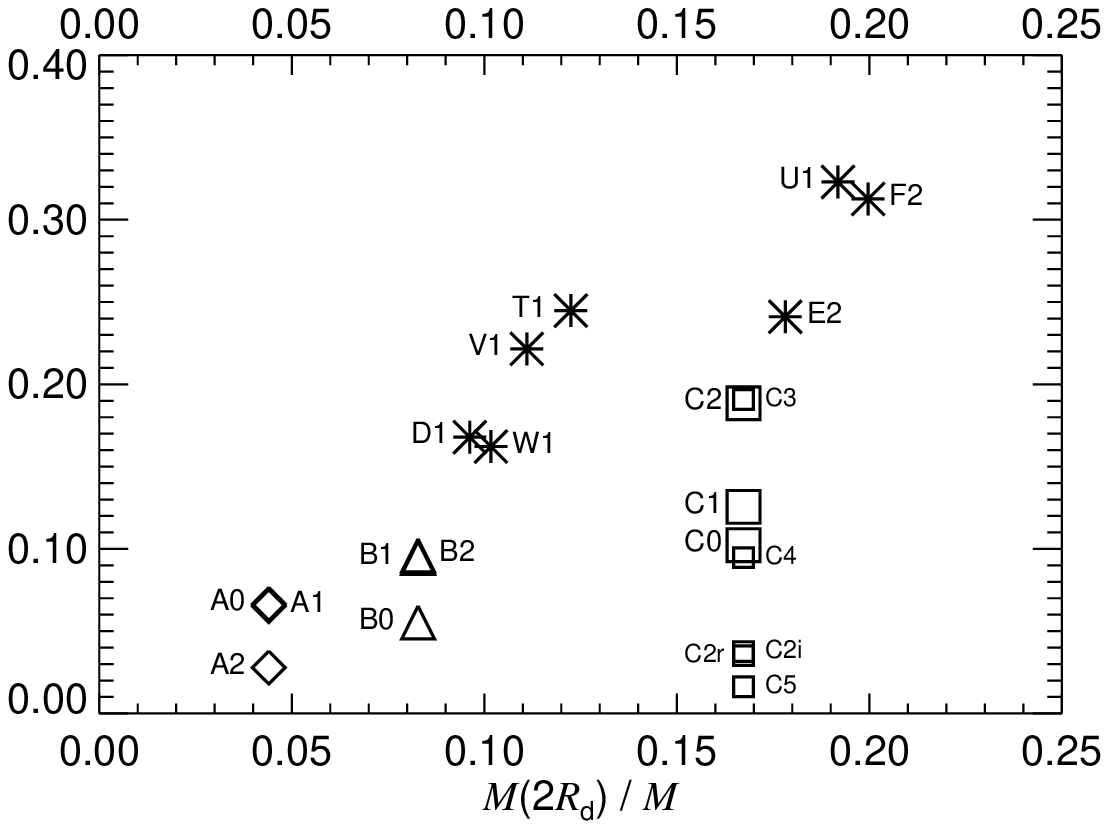}}\\
\caption{Tidal response $T_{\rm eff}$ 
of the different runs versus the disk-to-halo mass ratio (left panel),
and versus $M(2R_{\rm d})/M$ (right panel).
The quantity $M(2R_{\rm d})/M$ is a measure
of the fraction of the total mass in the disk region.
\label{response2}}
\ec
\end{figure*}

We now want to test how well ${\cal E}$ works as an indicator 
for the ability of 
a particular disk-halo model to develop massive tails.
Here one encounters two immediate problems. 

First, the tidal response of a disk depends  
on the orbital parameters of the encounter with its companion. For
example, if a disk is tilted against the orbital plane, the tidal
forces felt by the bulk of the disk material will generally be
smaller, resulting in a less prominent tidal tail. Similarly, a change
of the impact parameter or the orbital energy can affect the tidal
response. We here do not intend to investigate the complete parameter
space. Rather, we focus on collisions that produce the
strongest tails possible for mergers of a given disk model.

\begin{figure*}
\bc
\noindent{\includegraphics{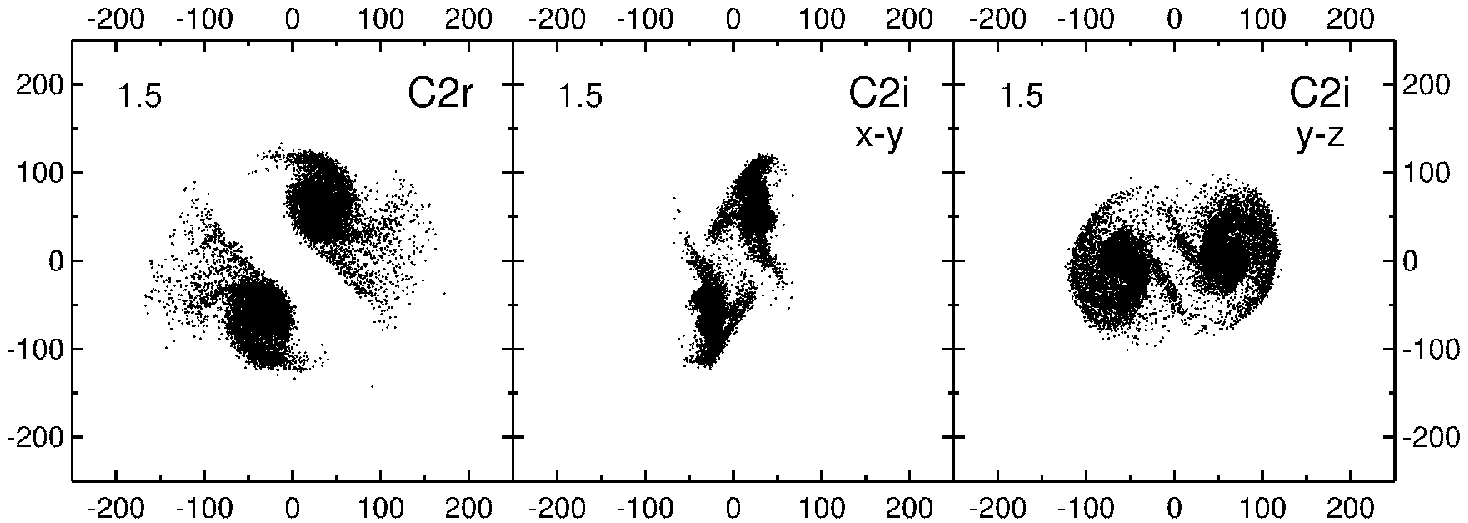}}\\
\caption{Tidal tails of simulations C2r and C2i.
The panels show the disk particles, projected either onto the xy-plane
(orbital plane), or the yz-plane.
Apart from a different orientation of the disks,
runs C2r and C2i are identical to C2. In C2r, both disks are
retrograde, while in C2i the spins of the disks lie in the orbital
plane, pointing along the x-axis.
These runs may be compared with C2 shown in Figure \ref{tailcomp1}.
\label{tailcomp3}}
\ec
\end{figure*}

In this spirit, prograde encounters are an obvious choice 
for the orientations of the disks, since this configuration has 
repeatedly been
shown to 
produce the strongest tails, and we here confirm this with the
runs C2r and C2i.
We use parabolic orbits, because they are plausible candidates
for real interacting galaxies if one assumes that 
they are coming together for their first time. 
However, even when this assumption is dropped, 
DMH showed that moderately bound 
orbits are no more effective in producing tails than zero energy orbits.
With respect to the `impact' parameter
$R_{\rm kep}$,
we have examined a
range of different choices and found that the tails 
appear to be of maximum strength for
$R_{\rm
min}\approx 2-6 R_{\rm d}$, i.e.\ in collisions where the disks pass
each 
other at a
distance of a few disk scale lengths.
Hence, our merger simulations of equal disk galaxies 
have been set up to exhibit 
the most favourable conditions for tail formation, and 
should indeed produce 
the strongest tails possible for these disk models.

Second, it is not obvious how to define the mass or extent of the tidal
tails in an objective way. This is further complicated by the rapid
dynamical evolution of the tails, which makes it difficult to compare
simulations that may form their tails at different times.

In order to solve this problem and measure the strength of the tidal response,
we have come up with the following scheme.
We start by defining the quantity
$T$ to be the mass fraction of each disk that reaches a
distance 
of more than $10R_{\rm d}$
from its center-of-mass, where
$R_{\rm d}$
is the original scale length of the unperturbed disk. 

In Figure \ref{resexamp} we show examples for the time evolution of
$T$. Shortly after the disks come together for the first
time, $T$ jumps up, reaches a maximum, and slowly decays, until the
disks are scrambled up in their second encounter and $T$ loses its
initial meaning.
Note that
the different runs reach their maximum of $T$ at different
times. In order to compare them on an equal footing, 
we therefore 
define an effective response $T_{\rm eff}$ as the peak value
reached by $T$.

In Figure \ref{response} 
we plot the tidal response $T_{\rm eff}$ of our runs
versus the value of ${\cal E}$, evaluated at $R=2R_{\rm d}$.
Although our coverage of parameter space is limited, it is
nevertheless clear that there is a correlation between 
$T_{\rm eff}$ and ${\cal E}$. 
Of course, there is some uncertainty in the measurement of
$T_{\rm eff}$ and this introduces some scatter. Also, for fixed ${\cal E}$,
the tidal
response $T_{\rm eff}$ depends somewhat on the impact parameter
$R_{\rm kep}$, and it might also have a slight dependence on $m_{\rm
d}$. 
However, to the extent that our simulations really
produce the strongest tails possible for our disk models, 
Fig.\ \ref{response} 
shows that it will be exceedingly hard to find a model
that produces tails that {\em lie above} the dashed diagonal line.
This establishes that ${\cal E}$ is a good indicator for the 
{\em maximum tidal response} $T_{\rm eff}$, 
that may be obtained for a given
class of disk models.
In particular, models with
${\cal E}\ge 8$ should be unable to produce strong tails. This is
in excellent agreement with the analysis of MMW, 
who estimated ${\cal E}=4.2$, 5.5, 7.2, and 9.3 (in
order of increasing halo mass) for
the sequence of four models of DMH; in agreement with Figure \ref{response},
the last two of these models failed to produce prominent tails.

Figure \ref{response} may also be compared to the two panels of 
Figure \ref{response2}. In the left panel, 
we plot the tidal response
versus the disk-to-halo mass ratio. This again shows, that the 
disk-to-halo mass ratio is not a good indicator for the ability to
form tidal tails. 
For example, the models A, B, and C differ
subtantially in the mass of their tails despite their equal disk-to-halo
mass ratio. 
However, if we use instead 
the ratio of the {\em total} amount of mass in the region of the disk 
to the total mass of
the galaxy, the mass ratio criterion can be partly resurrected. This is shown in the
right panel of Fig.\ \ref{response2}, where we plot $T_{\rm eff}$
versus $M(2R_{\rm d})/M$. Here $M(2R_{\rm d})$ is the total
mass inside two disk scale lengths, and $M$ is the total
mass of the galaxy and its halo. In this formulation,
the mass ratio measures the relative distribution of mass
within the system, and is a fair indicator of the ability of a galaxy model to
form tidal tails. However, a detailed comparison with Fig.\ \ref{response}
shows, that ${\cal E}$ does a better job than  $M(2R_{\rm
d})/M$. For example, the model T1 appears as an
outlyer in Fig.\ \ref{response2}, failing to fit 
the monotonic trend of larger $T_{\rm eff}$ with
increasing $M(2R_{\rm
d})/M$, but fits within the general distribution in Fig.\  \ref{response}.

Starting from a head on collision,
Figure \ref{response} also shows that the tidal response
$T_{\rm eff}$ becomes larger 
as the impact parameter $R_{\rm kep}$ is
increased.
However, for very wide encounters
one expects only a small distortion of the disks.
As the sequence of
models C0-C5 demonstrates, 
there is indeed a maximum response for an
intermediate 
impact parameter $R_{\rm kep}$ of order a few disk scale lengths. 

Also, the orientation of the disks is an important factor 
in determining the strength of the 
disk response. In the retrograde collision C2r, the
spins of the galaxies are just reversed compared to C2, yet this
already 
makes the tidal tails much weaker, as seen in Figure \ref{tailcomp3}. 
Similarly, the inclined galaxies of
simulation C2i produce tails that are less extended than those
of run C2.

We also note that the simulations T1, U1, and V1, 
in which our 
`disk-dominated' models collide, produce tidal tails well in
line with the general trend of Figure \ref{response}.

\subsection{A model with a bulge}

\begin{figure*}
\bc
{\includegraphics{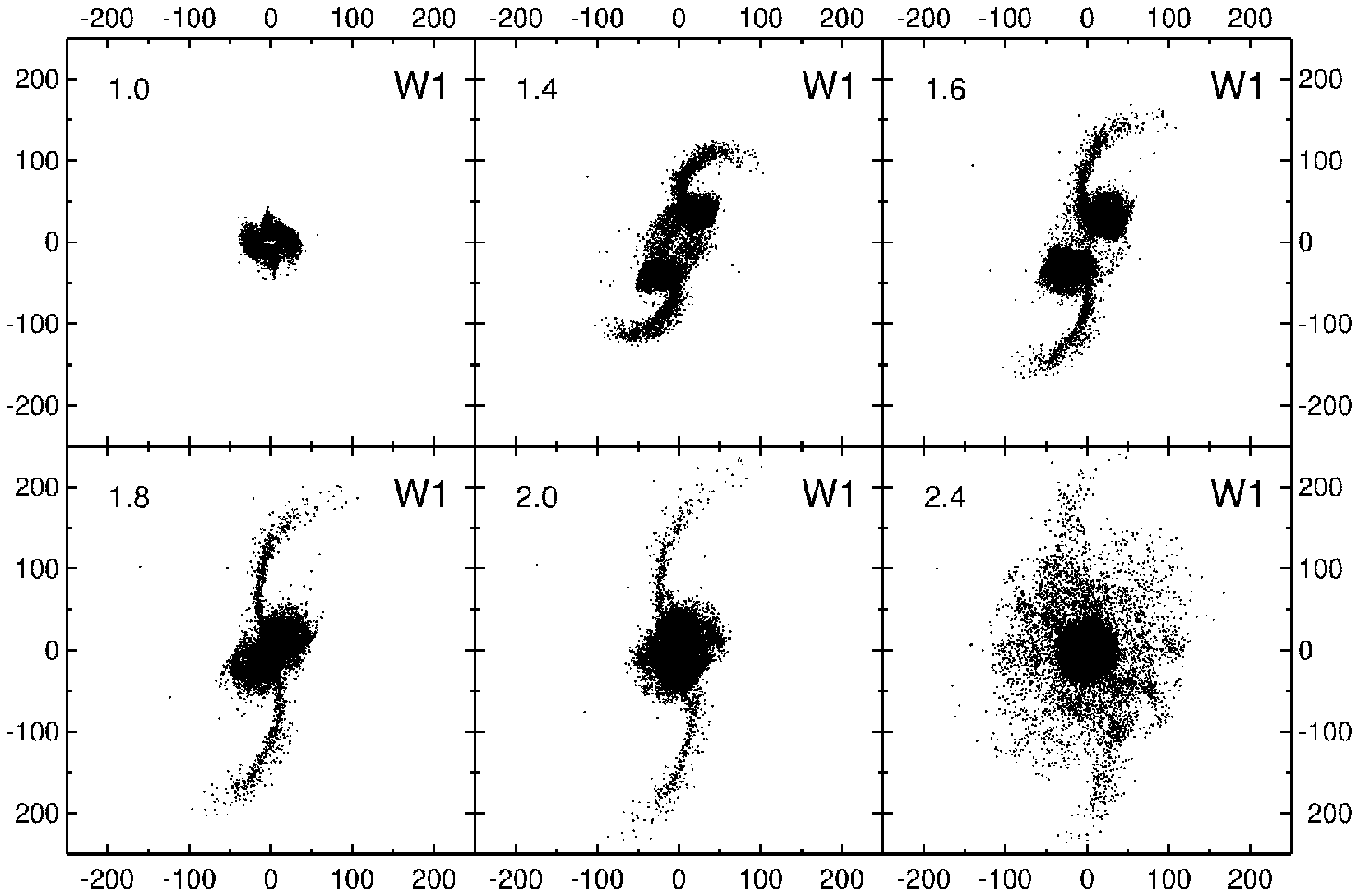}}\\
{\includegraphics{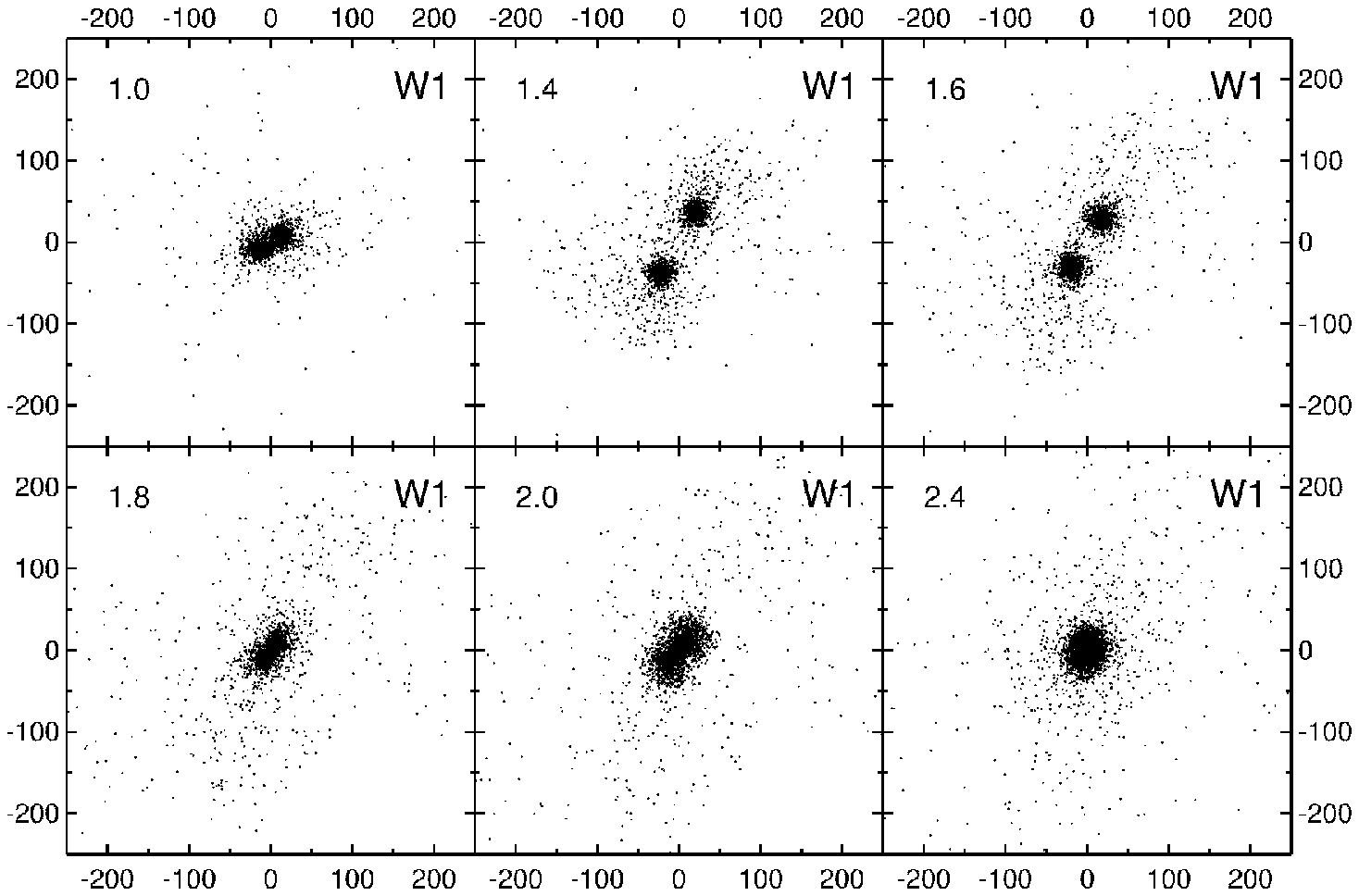}}\\
\caption{Time evolution of run W1, which collides our disk+bulge model
with itself.
The top series of panels 
displays the disk particles projected onto the orbital
plane, while the bottom panels show the bulge particles.
 The length units refer to $\lu$,
and the elapsed time since the start of the simulation (upper left
corners)
is given in units of 0.1 Hubble times, or $9.8\times 10^8\,h^{-1}{\rm yr}$.
\label{evolvBulge}}
\ec
\end{figure*}

In Figure \ref{evolvBulge} we show the time evolution of run W1,
which collides two disk$\,+\,$bulge galaxies.
These disk galaxies (our model W) are descendents of model D, but one
third of the stellar mass has been put into a centrally concentrated bulge.
This results in a rotation curve (Figure \ref{rotcurvebulge}) 
that is practically flat up to the
very center of the disk, with a shape quite similar to the models of
DMH.

Due to their strong central concentration, the bulges survive largely
unaffected until their final coalescence. However, the disks develop
prominent tails that appear to be similar in strength to the other
models with specific angular momentum corresponding to $\lambda=0.05$.
A measurement of $T_{\rm eff}$ shows that the strength of the
tails is in fact
quite similar to the directly comparable simulation D1.
Also, the run W1 fits well into the plot of Figure \ref{response},
although the inner structure of model W is very different from that of
the other models.

This suggests that bulges are not effective in preventing tail
formation, 
at least as long as they primarily affect the inner rotation curve.

\section{Discussion}

In this study we constructed N-body models of disk galaxies with
structural properties directly motivated by current theories of hierarchical
galaxy formation. In particular, the mass of the dark haloes in these
models is much larger than that of the stellar disks. In the most
extreme models we consider, the halo-to-disk mass ratio is larger than 40:1.

Provided the spin parameter is not too small,
these models can produce
long and massive tidal tails, despite the massive haloes.
Halo-to-disk mass ratio
is not a useful indicator for tail-making
ability. 

Instead, the size of the disk compared to that of the halo seems to be
the critical factor. In our approach, the size of the disk is tied
to the spin of the dark halo. 
A larger spin parameter $\lambda$ leads to larger disks. The bulk
of the disk material is then more loosely bound in the 
dark matter potential well and can be more easily induced to form long tidal
tails.
This effect can be quantified in terms
of
the ratio ${\cal E}$ of the circular velocity 
to the escape speed at a radius $R=2 R_{\rm d}$. 
We have shown that ${\cal E}$ correlates well with the tidal response
of the disk models. For models with ${\cal E}\ge 8$ we do not expect
significant tails, while models with ${\cal E}\le 6.5$ can
produce substantial tails.

When ${\cal E}$ is used to
characterize the tail-making ability,
the results of DMH agree with our own.
In their sequence of four models 
not only the mass of the halo changes, but also its spatial extent. We think 
the latter effect is critical in defining
tail-making ability, 
because the relative size of disk and halo 
affects
the value of ${\cal E}$
strongly.
This also agrees with the earlier conclusion of Barnes (1997, private
communication). For given {\em inner} structure and given mass ratio 
$M_{\rm d}/M_{\rm
h}$, less extensive halos have larger ${\cal E}$ and so make weaker tails.

We focused in this work on just one halo mass. Note however, that the
shape of the rotation curves does not depend on our particular choice for
$v_{200}$. Rotation curves with other peak velocities may be
realized by an appropriate scaling of $v_{200}$.

According to the work of NFW, the shape of the dark matter profile is
insensitive to cosmology. Also, the distribution of $\lambda$ is
universal and independent of the initial power spectrum. 
Furthermore, the average value of the concentration 
$c$ does not vary with halo mass,
and it depends only weakly on cosmology. For low-density flat
universes a smaller value, $c\simeq 5$, is probably more
appropriate than the value $c=15$ employed in most of the models in this work.
Such a smaller concentration is also supported by observational data \citep{Na98}.
However, our model `V1', which has $c=5$, produces tails well in line
with the other models in Figure \ref{response}. 
Also note that according to Figure \ref{rdlambda}, smaller $c$ 
gives rise to larger disks, in principle favouring even stronger
tails.

This suggests
that all `reasonable' CDM cosmologies can produce disk galaxies with 
$\lambda\ge 0.05$ that
are roughly equally capable of producing tidal tails
when they collide and merge with similar objects. 
We conclude that the observed lengths of tidal tails in interacting
galaxies are consistent with current CDM cosmologies, and that tidal
tails are not useful  
to discriminate between different flavours of
these scenarios.

We hope that the N-body representations of disk models constructed in
this work may be more realistic caricatures of real spiral galaxies than
those of previous work. In particular, the structural properties of
our models are motivated by hierarchical structure formation and are 
less ad hoc than in
previous simulations of this kind. 
These models should be useful for future work on galaxy evolution.

\section*{Acknowledgements}

SW would like to thank 
Joshua Barnes for useful discussions about tail making 
and to note that he saw preliminary results from Barnes' own
simulations in July 1997, six months before the current project was
started.

\bibliographystyle{mnras}
\bibliography{paper}

\end{document}